\documentclass[10pt]{article}

\usepackage[utf8]{inputenc}
\usepackage[T1]{fontenc}

\usepackage{epsf}
\usepackage{amsmath}

\allowdisplaybreaks

\usepackage[showframe=false]{geometry}
\usepackage{changepage}

\usepackage{epsfig}
\usepackage{amssymb}

\usepackage{amsthm}
\usepackage{setspace}
\usepackage{cite}
\usepackage{mcite}

\usepackage{algorithmic}  
\usepackage{algorithm}

\usepackage{shadow}
\usepackage{fancybox}
\usepackage{fancyhdr}

\usepackage{color}
\usepackage[usenames,dvipsnames,svgnames,table]{xcolor}
\newcommand{\bl}[1]{\textcolor{blue}{#1}}

\definecolor{mypurple}{rgb}{.4,.0,.5}

\usepackage[hyphens]{url}

\usepackage[colorlinks=true,
            linkcolor=black,
            urlcolor=black,
            citecolor=purple]{hyperref}

\usepackage{breakurl}

\def\y{{\bf y}}

\def\x{{\bf x}}

\def\x{{\mathbf x}}

\def\u{{\bf u}}

\def\x{{\bf x}}
\def\y{{\bf y}}

\def\q{{\bf q}}

\def\a{{\bf a}}
\def\b{{\bf b}}
\def\c{{\bf c}}

\def\h{{\bf h}}

\def\tr{\mbox{Tr}}

\def\tr{{\rm tr}\,}

\def\cH{{\mathcal H}}

\def\be{\begin{equation}}
\def\ee{\end{equation}}
\def\ba{\left[\begin{array}}
\def\ea{\end{array}\right]}

\def\u{{\bf u}}

\def\x{{\bf x}}
\def\y{{\bf y}}

\def\q{{\bf q}}
\def\a{{\bf a}}
\def\b{{\bf b}}
\def\c{{\bf c}}

\def\p{{\bf p}}

\def\1{{\bf 1}}

\def\g{{\bf g}}
\def\0{{\bf 0}}

\def\erfc{\mbox{erfc}}

\def\cG{{\mathcal G}}

\def\mR{{\mathbb R}}

\def\mN{{\mathbb N}}
\def\mE{{\mathbb E}}
\def\mS{{\mathbb S}}
\def\mB{{\mathbb B}}
\def\mP{{\mathbb P}}

\def\lp{\left (}
\def\rp{\right )}

\def\y{{\bf y}}

\def\x{{\bf x}}

\def\x{{\mathbf x}}

\def\u{{\bf u}}

\def\x{{\bf x}}
\def\y{{\bf y}}

\def\q{{\bf q}}
\def\a{{\bf a}}
\def\b{{\bf b}}
\def\c{{\bf c}}

\def\h{{\bf h}}

\def\tr{\mbox{Tr}}

\def\tr{{\rm tr}\,}

\def\cH{{\cal H}}

\def\be{\begin{equation}}
\def\ee{\end{equation}}
\def\ba{\left[\begin{array}}
\def\ea{\end{array}\right]}

\def\u{{\bf u}}

\def\x{{\bf x}}
\def\y{{\bf y}}

\def\q{{\bf q}}
\def\a{{\bf a}}
\def\b{{\bf b}}
\def\c{{\bf c}}

\def\p{{\bf p}}

\def\({\left (}
\def\){\right )}

\def\1{{\bf 1}}

\def\q{{\bf q}}

\def\g{{\bf g}}
\def\0{{\bf 0}}

\def\cX{{\mathcal X}}
\def\cY{{\mathcal Y}}
\def\cS{{\mathcal S}}

\def\cA{{\mathcal A}}

\def\cX{{\mathcal X}}

\usepackage{xcolor}
\usepackage{color}

\definecolor{darkgreen}{rgb}{0, 0.4,0}

\definecolor{purplebrown}{rgb}{0.5,0.1,0.6}

\definecolor{ultclupcol}{rgb}{0.1,0.5,0.5}

\definecolor{mytrycolor}{rgb}{0.5,0.7,0.2}

\definecolor{ultclupcola}{rgb}{.5,0,.5}

\definecolor{shadebrown}{rgb}{0.1,0.1,0.9}
\definecolor{lightblue}{rgb}{0.2,0,1}

\usepackage{fancybox}
\usepackage{graphicx}
\usepackage{epstopdf}
\usepackage{epsfig}
\usepackage{wrapfig}
\usepackage{subfigure}

\usepackage{xcolor}
\usepackage{tcolorbox}

\newtcbox{\xmybox}{on line,
arc=7pt,
before upper={\rule[-3pt]{0pt}{10pt}},boxrule=0pt,
boxsep=0pt,left=6pt,right=6pt,top=0pt,bottom=0pt,enhanced, coltext=blue, colback=white!10!yellow}

\newtcbox{\xmyboxa}{on line,
arc=7pt,
before upper={\rule[-3pt]{0pt}{10pt}},boxrule=0pt,
boxsep=0pt,left=6pt,right=6pt,top=0pt,bottom=0pt,enhanced, colback=white!10!yellow}

\newtcbox{\xmyboxb}{on line,
arc=7pt,
before upper={\rule[-3pt]{0pt}{10pt}},boxrule=1pt,colframe=darkgreen!100!blue,
boxsep=0pt,left=6pt,right=6pt,top=0pt,bottom=0pt,enhanced, colback=white!10!yellow}

\newtcbox{\xmyboxc}{on line,
arc=7pt,
before upper={\rule[-3pt]{0pt}{10pt}},boxrule=.7pt,colframe=blue!100!blue,
boxsep=0pt,left=6pt,right=6pt,top=0pt,bottom=0pt,enhanced, coltext=blue, colback=white!10!yellow}

\newtcbox{\xmytboxa}{on line,
arc=7pt,
before upper={\rule[-3pt]{0pt}{10pt}},boxrule=.0pt,colframe=pink!50!yellow,
boxsep=0pt,left=6pt,right=6pt,top=0pt,bottom=0pt,enhanced, coltext=white, colback=blue!40!red}

\newtcbox{\xmytboxb}{on line,
arc=7pt,
before upper={\rule[-3pt]{0pt}{10pt}},boxrule=.0pt,colframe=pink!50!yellow,
boxsep=0pt,left=6pt,right=6pt,top=0pt,bottom=0pt,enhanced, coltext=white, colback=white!40!green}

\makeatletter
\newcommand\subsubsubsection{\@startsection{paragraph}{4}{\z@}{-2.5ex\@plus -1ex \@minus -.25ex}{1.25ex \@plus .25ex}{\normalfont\normalsize\bfseries}}
\newcommand\subsubsubsubsection{\@startsection{subparagraph}{5}{\z@}{-2.5ex\@plus -1ex \@minus -.25ex}{1.25ex \@plus .25ex}{\normalfont\normalsize\bfseries}}
\makeatother

\newtheorem{theorem}{Theorem}

\newtheorem{corollary}{Corollary}

\begin{document}

\begin{singlespace}

\title {Random matrix discrepancy -- CLuP algorithm and parametric RDT approach
}
\author{
\textsc{Mihailo Stojnic
\footnote{e-mail: {\tt flatoyer@gmail.com}}
}}
\date{}
\maketitle

\centerline{{\bf Abstract}} \vspace*{0.1in}

We study information-theoretic and algorithmic aspects of random matrix discrepancy in proportional regime. The problem amounts to finding $\kappa=\min_{\x\in\mB^n}\|\sum_{i=1}^{n}\bar{A}_i\x_i\|_2$, where $\bar{A}_i\in\mR^{d\times d}$ is a collection of $n$ iid GOE matrices, $\mB^n$ are vertices of the unit cube in $\mR^n$, and $\alpha=\lim_{n\rightarrow \infty}\frac{n}{d^2}$ is a proportionality constant. 

Strong progress is made in several directions: (i) We develop a Random duality theory (RDT) framework and utilize its a lifted variant to establish lower bounds on $\kappa$, which precisely match the bounds obtained via the first moment in \cite{Maill24}. (ii) To address analytical challenges, we propose a negative Hopfield model as a proxy for implementation of a fully lifted RDT. On the 2nd lifting level, this model fits remarkably well between the information-theoretic lower bounds from \cite{Maill24} and the algorithmic upper bounds from \cite{ChenAlaoui26}. (iii) On the 3rd and 4th lifting levels, parametric RDT concepts predict a sizeable computational gap for lower $\kappa$ values in the proxy model. Extending proxy information-theoretic fitting to algorithmic suggests a similar computational gap for matrix discrepancy in low $\kappa$ regime, aligning with \cite{KizLi26}. (iv) We design a Controlled Loosening Up (CLuP) algorithm for matrix discrepancy. Its performance strongly matches our analytical results -- closely approaching information-theoretic predictions in lower $\alpha$ ranges (indicating a small or nonexistent computational gap) and following the proxy model's algorithmic predictions in higher $\alpha$ ranges  (indicating a sizeable computational gap).

Both the information-theoretic and algorithmic predictions match the corresponding phenomenologies of the symmetric binary perceptron (SBP) vector analogue in the entire $\alpha\in(0,\infty)$ range \cite{AbbLiSly21b,PerkXu21,AbbLiSly21a,AubPerZde19,GamKizPerXu22,BarbAKZ23,Stojnicalgbp25}.

\vspace*{0.25in} \noindent {\bf Index Terms: Matrix discrepancy; CLuP algorithm; Random duality theory (RDT)}.

\end{singlespace}

\section{Introduction}
\label{sec:back}

Tremendous AI progress over the last decade has brought perceptrons back into the spotlight as key machine learning units. Objects studied in classical works \cite{Wendel62,Winder,Cover65} over 60 years ago have become relevant again.

The horizons quickly widened, and many perceptron variants -- including positive \cite{Gar88,GarDer88,SchTir02,SchTir03,Tal05,Talbook11a,Talbook11b,StojnicGardGen13} and negative spherical ones \cite{FPSUZ17,FraHwaUrb19,FraPar16,FraSclUrb19,FraSclUrb20,AlaSel20,BMPZ23}, as well as asymmetric \cite{KraMez89,DingSun19,NakSun23,BoltNakSunXu22,Huang24,Stojnicalgbp25} and symmetric binary ones \cite{AbbLiSly21b,PerkXu21,AbbLiSly21a,AubPerZde19,GamKizPerXu22,Stojnicalgsbp26} -- have been thoroughly reexamined. They often assume key roles in neural network (NN) and machine learning (ML) developments, either as integral architectural components or as simple prototype models used to conceptually emulate more complex structures.

Symmetric binary perceptrons (SBP) \cite{AbbLiSly21b,PerkXu21,AbbLiSly21a,AubPerZde19,GamKizPerXu22,Stojnicalgsbp26} have received particular attention, as their structural simplicity offers key advantages regarding analytical tractability. In particular, they allow for the simultaneous study of storage capacities (as a key neural network feature) and statistical-computational gaps (SCG, as key computational barriers). SBPs offer a rare combination of approachable analytical tractability and presumed unsurpassable algorithmic difficulty.

Widely known as feasibility problems in ML communities, SBPs also have a dual $\infty$-norm minimization formulation known as minimum vector balancing or minimum vector discrepancy. Consequently, they have been intensively studied in the discrete mathematics and theoretical computer science communities for nearly half a century \cite{Spen85,Chazelle00,Harshaw24,BanSpen20,Matousek09,ChenSriTra2014,Costello09,TuMR20,KarKLO86,Levy16,BanSpen11}.

In this work, we are particularly interested in matrix discrepancy (MD) problems, which are matrix variants of vector balancing or SBPs \cite{Zouzias12,BanJiaMeka23,HopRS22,kathuria26a,Dadush22,kathuria26b,MarSS15,BatsonSS14,BandBol26,AkbSra26,KunOWZ24,Maill24}. Motivated in part by the nature of machine learning data, recent studies of SBP and vector discrepancy problems have focused on random mediums. Continuing this trend, we consider random mediums to study both the information-theoretic and algorithmic aspects of MD problems.

We primarily focus on precise analyses capable of pinpointing phase transition locations and the potential emergence of statistical-computational gaps (SCG). Our main findings indicate that SCG phenomenologies present in SBP and negative Hopfield models also appear in matrix discrepancy: (i) For smaller matrix sizes (larger $\alpha$): Problems are computationally difficult, featuring a sizeable SCG. (ii) For larger matrix sizes (smaller $\alpha$): Problems are either computationally tractable or accessible via algorithms that closely approach information-theoretic limits.

Before stating and positioning our results in detail, we briefly review recent relevant literature.

\subsection{ Relevant prior work }
\label{secanaldiff}

\noindent{\underline{\textbf{\emph{Vector discrepancy}}}}  Studying discrepancy problems typically starts with vector variants. For a given matrix $A\in\mR^{m\times n}$ with elements not exceeding 1, vector discrepancy asks for characterization of
\begin{equation}\label{eq:inteq1}
\min_{\x\in\mB^n}\|A \x\|_\infty
\end{equation}
where $\mB^n\triangleq\left \{-\frac{1}{\sqrt{n}} ,\frac{1}{\sqrt{n}}\right \}^n  $. This problem has a rich history spanning a variety of scientific fields with a diverse range of applications (see, e.g., \cite{Barany81,EisenWeis18,BansalCKL14,HobRoth17,ChandraVemp14,JanRoh22,Nikolov17}). One of the major breakthroughs in this area is Spencer's upper bound of $O(n\log(2m/n))$ \cite{Spen85} (see also \cite{BeckFiala81}). While Spencer's result was non-constructive, it was eventually complemented by corresponding algorithmic approaches \cite{BanSpen11,Bansal10,LovettMeka15}. For further algorithmic developments along these lines, see, e.g., \cite{Levy16,Rothvoss17} and references therein. For recent progress and resolutions regarding the Komlos variant of the problem--where the columns of $A$ are restricted to have a Euclidean norm (instead of an $\infty$-norm) smaller than 1--see \cite{BanJia26,BanJia26b,KarLov26,GuoFL26}. For sparse vectors and additional results on the Beck-Fiala and Komlos conjectures, see, e.g., \cite{Banaszczyk98,Banaszczyk12,Gluskin88,Giann97}.

A key challenge in discrepancy theory is whether a random choice of $\pm$ signs can be beaten, which Spencer's result and subsequent algorithmic realizations successfully achieve. Specifically, for square systems ($n=m$), a naive random choice of signs yields only $O(\sqrt{n\log(n)})$, whereas \cite{Spen85} achieves $O(\sqrt{n})$.

In the random context, the classical example is the standard Gaussian, where $A$ is comprised of independent standard normals (for different types of randomness, see, e.g., \cite{AltNilesWeed22,HobRoss19}). For small loadings $m=o(n)$, discrepancy behaves roughly as $O\lp \sqrt{n} 2^{-\frac{n}{m}}\rp$ \cite{TuMR20}, while no known polynomial algorithm achieves better than $2^{-O(\log^2\lp n/m\rp)}$ \cite{Costello09,TuMR20,KarKLO86} (for closely related number partitioning considerations where $m=1$, see also \cite{Vafa25,GamKiz23}).

In the linear (proportional) regime $\alpha_{SBP}=\lim_{n\rightarrow \infty }\frac{m}{n}$, the precise values of achievable (so-called information-theoretic) discrepancy $\kappa$ have been determined in \cite{AbbLiSly21b,PerkXu21,AbbLiSly21a,AubPerZde19}. It is predicted that the problem exhibits a fairly large SCG, particularly in the low $\kappa$ regime. Namely, for small $\kappa$, the information-theoretic estimate is $\alpha_{SBP}=O\lp \frac{1}{-\log(\kappa)}\rp$ \cite{AbbLiSly21b,PerkXu21,AbbLiSly21a,AubPerZde19}, whereas the corresponding algorithmic one is believed to be far away $\alpha_{SBP}=O(-\kappa^2\log(\kappa))$ \cite{GamKizPerXu22,BarbAKZ23,Stojnicalgbp25} (modulo the $\log$ term, it is actually achievable by known polynomial algorithms \cite{BanSpen20}).

\vspace{.1in}

\noindent{\underline{\textbf{\emph{Matrix discrepancy}}}}  Given an array of matrices, $\cA=\{\bar{A}_i\}_{i=1,2,\dots,n}$, where $\bar{A}_i=\bar{A}_i^T\in\mR^{d\times d}$ and $\|A_i\|_2\leq 1$, the following  is mathematical formulation of the matrix discrepancy problem 
\begin{equation}\label{eq:inteq2}
\min_{\x\in\mB^n} \left \|\sum_{i=1}^{m} \bar{A}_i \x_i \right \|_\infty.
\end{equation}
These problems have also been utilized in various further studies (see, e.g., \cite{MarSpiSri15,HopRS22,ReisRoth20}, where applications in graph theory, quantum systems, and the Kadison-Singer problem are discussed).

The vector variant from (\ref{eq:inteq1}) is recovered for diagonal $\bar{A}_i$. Logic similar to the one employed for the vector variant applies here as well. Namely, for a random choice of signs, one has discrepancy $O(\sqrt{n\log(d)})$, and one might wonder whether such a choice can be beaten \cite{Zouzias12}. Significant progress has been made over the years while studying various special cases \cite{BanJiaMeka23,HopRS22,Dadush22,BandBol26,MarSS15,BatsonSS14}, with the most general of them being \cite{BanJiaMeka23}, where just a tiny rank constraint $O\lp\frac{n}{\log^3(d)}\rp$ on the matrices is assumed. Very recently, however, a series of excellent works has completely settled the problem \cite{AkbSra26,AkbSra26b,AkbSra26c,BanKook26}. Moreover, the corresponding algorithmic results have been obtained as well \cite{SongZhang26,kathuria26a,kathuria26b}.

Studying random matrix discrepancy (the main topic of our paper) was initiated in \cite{KZ23} (see also \cite{KunOWZ24,Maill24}). Considering again the classical Gaussian setting (symmetric GOE matrices), \cite{KunOWZ24} showed that for $d=o(\sqrt{n})$, the matrix discrepancy $\kappa$ is $O\lp 4^{-\alpha}\rp$. Under additional assumptions, \cite{KunOWZ24} further extends these results to other types of randomness and considers complementary algorithmic results, focusing on online algorithms.

In \cite{Maill24}, the linear (proportional) regime with $\alpha=\lim_{n\rightarrow \infty}\frac{n}{d^2}$ was considered (note that this notation may be viewed as inverted compared to SBP, i.e., $\alpha\sim \alpha_{SBP}^{-1}$). Utilizing the first moment approach and large deviations of particularly correlated GOE ensembles, \cite{Maill24} determined the lower and upper bounds on the discrepancy $\kappa$ throughout the entire range of $\alpha$. Interestingly, to the leading order as $\kappa\rightarrow 0$, the results of \cite{Maill24} match the $O(4^{-\alpha})$ sublinear regime estimate from \cite{KunOWZ24} (\cite{Maill24} also provides a corresponding constant, which naturally differs in the proportional regime).

Very recent works \cite{KizLi26} and \cite{ChenAlaoui26} consider complementary algorithmic aspects in the proportional setting. In \cite{ChenAlaoui26}, a variant of I-AMP (incremental approximate message passing) was designed, inspired by the full-RSB algorithmic approach to spherical SK models \cite{Subag21} (I-AMP was originally introduced for solving the Ising SK model in \cite{Montanari19}; see also \cite{HuangS22,AlaouiMS21} for further progress). Through a particular choice of I-AMP-associated nonlinearities, \cite{ChenAlaoui26} obtained an explicit algorithmic upper bound on matrix discrepancy $\kappa$ for a large range of $\alpha$ (the so-called lower $\alpha$ regime or higher $\kappa$ regime).

On the other hand, \cite{KizLi26} focused primarily on the lower $\kappa$ (higher $\alpha$) regime and showed that no stable algorithms can achieve discrepancy smaller than $O\lp\frac{1}{\sqrt{\alpha\log(\alpha)}}\rp$. That is, for small $\kappa$, polynomial stable algorithms require that the loading $\alpha$ is at least $O\lp \frac{1}{-\kappa^2\log(\kappa)} \rp$. Additionally, an analogous $\alpha$ limitation of $O\lp \frac{1}{\kappa^2} \rp$ is obtained for online algorithms in \cite{KizLi26} as well. Moreover, \cite{KizLi26} matches both of these results with concrete algorithms: a \emph{recenter-and-round} algorithm for offline settings and a \emph{Frobenius-greedy} algorithm for online settings.

Remarkably, keeping in mind the inverted notational scaling $\alpha\sim\alpha_{SBP}^{-1}$, the standard offline results precisely match the SBP ones obtained in \cite{GamKizPerXu22,BarbAKZ23,Stojnicalgbp25}. Adding to this that Maillard's lower bound on $\alpha$ in the small $\kappa$ regime behaves as $\sim -\log(\kappa)$ -- which, with the notational inversion $\alpha\sim\alpha_{SBP}^{-1}$, matches the SBP behavior from \cite{AbbLiSly21b,PerkXu21,AbbLiSly21a,AubPerZde19} -- the results of \cite{Maill24} and \cite{KizLi26} suggest that SBP and matrix discrepancy may have similar statistical-computational gaps, at least in the low $\kappa$ range.

\subsection{Our contributions}
\label{sec:cont}

Below is the summary of our key findings regarding the information theoretic and algorithmic aspects of random matrix discrepancy in the proportional regime across the entire $\alpha \in (0, \infty)$ range:

\begin{itemize}
\item RDT framework: We developed the Random Duality Theory (RDT) framework for studying matrix discrepancy. Utilizing a lifted RDT variant, we proved lower bounds on the information theoretic discrepancy values $\kappa$, which precisely match the lower bounds obtained via the first moment in \cite{Maill24}. (See Section \ref{sec:implem}.)

\item Proxy model: As the problem that we study is extremely hard, a direct attack via fully lifted RDT (analogous to what was done for SBP in \cite{Stojnicalgsbp26} and for ABP in \cite{Stojnicalgbp25}) is likely to face substantial computational hurdles. As an alternative, we introduce a negative Hopfield proxy model to study algorithmic aspects and analyze it via fully lifted RDT. (See Section \ref{sec:proxy1}.)

\item Parametric RDT Predictions: Concepts from \cite{Stojnicalgbp25,Stojnicalgsbp26} predict for proxy that a computational gap is present in higher $\alpha$ ranges and absent in lower ranges. In lower $\alpha$ ranges, 4th-level lifting predictions are within $\sim 0.001$ of the algorithmic results from \cite{ChenAlaoui26}, suggesting the gap is indeed small or non-existent. Furthermore, our proxy information theoretic predictions remain above the information theoretic lower bounds from \cite{Maill24} and Section \ref{sec:implem} across all $\alpha \in (0, \infty)$. (See Section \ref{sec:implemh}, Section \ref{sec:disc1}, Table \ref{tab:tab1}, and Figure \ref{fig:fig1}.)

\item Algorithmic Trends: Using 3rd and 4th levels of lifting, we observed that proxy exhibits a sizeable computational gap in the higher $\alpha$ range. Assuming the proxy model reflects the intrinsic features of matrix discrepancy, a similar computational gap is expected for matrix discrepancy in higher $\alpha$ regimes. (See Sections \ref{sec:rthlev} and \ref{sec:disc1}, and Figure \ref{fig:fig1}.)

\item CLuP Algorithm: To practically solve matrix discrepancy, we designed a CLuP (controlled loosening up) algorithm. Its performance aligns remarkably well with analytical predictions already for dimensions on the order of $\sim 1000$. It closely approaches information theoretic predictions in lower $\alpha$ ranges and follows the proxy model's algorithmic predictions in higher $\alpha$ ranges, signaling the presence of a computational gap at higher $\alpha$. (See Section \ref{sec:algimp} and Figure \ref{fig:fig1}.)
\end{itemize}

In summary, these results suggest the following picture for random matrix discrepancy problems in the proportional regime:
\begin{itemize}
\item \emph{Information theoretic $\kappa$ thresholds}: The lower bounds obtained via first moment analysis in \cite{Maill24} and via lifting in Section \ref{sec:implem} are likely very close to the true values across the entire $\alpha$ range.

\item \emph{Algorithmic $\kappa$ thresholds}: In lower $\alpha$ ranges, the computational gap is small or non-existent; in higher $\alpha$ regimes, a significant computational gap appears. This aligns with recent results from \cite{KizLi26} showing stable offline algorithms achieve discrepancy no less than $O\lp\sqrt{\frac{1}{\alpha\log(\alpha)}}\rp$ (which is far away from information theoretic $O(4^{-\alpha})$ bound).

\item \emph{Analogy with SBP}: While both matrix discrepancy and SBPs exhibit computational gaps that shrink as $\kappa$ increases, the shrinking is more pronounced in matrix discrepancy, opening the possibility of a zero computational gap in certain low $\alpha$ regimes.
\end{itemize}

\section{Mathematical setup formalization}
 \label{sec:bprfps}

\subsection{Optimization and free energy problem formulations}
 \label{sec:bprfps}

Let $m$ and $n$ be two positive integers such that $\alpha_{SBP}=\frac{m}{n}$.  We start things off by recalling on classical \emph{vector}  discrepancy minimization problems. For matrix $A\in\mR^{m\times n}$ (see, e.g., \cite{KaKLO86,Spen85,LovMek15,GamKizPerXu23,Roth17,AlwLiuSaw21}), one considers the following optimization program
\begin{eqnarray}
\hspace{-1.5in}\mbox{$\mathbf{\mathcal D}(G,\alpha)$:} \hspace{1in}\min_{\x} & & \|A\x \|_{\infty} \nonumber \\
\mbox{subject to}
 & & \x\in \left \{-\frac{1}{\sqrt{n}},\frac{1}{\sqrt{n}} \right \}^n \triangleq \mB^n. \label{eq:ex1a1}
\end{eqnarray}
Practically speaking, the above optimization is trying to find a (min-max) balanced split of $n$ $m$-dimensional vectors. The optimization objective evaluates how far off from a perfect balance (zero objective) one is.  The following \emph{feasibility} alternative of (\ref{eq:ex1a1}) is also well known 
\begin{eqnarray}
\hspace{-1.5in}\mbox{$\mathbf{\mathcal S}(G,\b,\alpha)$:} \hspace{1in}\mbox{find} & & \x\nonumber \\
\mbox{subject to}
& & \|A\x\|_{\infty}\leq \kappa \1 \nonumber \\
& & \x\in \left \{-\frac{1}{\sqrt{n}},\frac{1}{\sqrt{n}} \right \}^n. \label{eq:ex1a0a0}
\end{eqnarray}
Within the neural network, machine learning, and statistical physics communities, the above alternative is typically referred to as the symmetric binary perceptron (SBP) \cite{Stojnicalgsbp26,AubPerZde19,AbbLiSly21a,AbbLiSly21b,Bald20,GamKizPerXu22,PerkXu21,ElAlGam24,SahSaw23,Barb24,djalt22,BarbAKZ23}. For the classical asymmetric binary perceptron (ABP) variant obtained by removing the $\|\cdot\|_{\infty}$, see \cite{Stojnicalgbp25,Talbook11a,StojnicGardGen13,GarDer88,Gar88,KraMez89,GutSte90,KimRoc98,NakSun23,BoltNakSunXu22,PerkXu21,CXu21,DingSun19,Huang24,LiSZ24}. Changing the $\x$ domain further to the unit sphere yields positive spherical perceptrons for $\kappa\geq 0$ \cite{StojnicGardGen13,GarDer88,Gar88,Schlafli,Cover65,Winder,Winder61,Wendel62,Cameron60,Joseph60,BalVen87,Ven86,SchTir02,SchTir03} and negative spherical perceptrons for $\kappa\leq 0$ \cite{AMZ24,BMPZ23,FPSUZ17,Talbook11a,FraHwaUrb19,FraPar16,FraSclUrb19,FraSclUrb20,AlaSel20}.

As noted above, excellent information-theoretic and algorithmic results have been obtained over the last several years regarding all of these perceptron versions. Particularly strong progress has been made regarding random (or average-case) instances, which are also the primary focus of our paper. However, instead of classical perceptrons, we are interested in matrix discrepancy problems, which are matrix generalizations of the vector discrepancy (\ref{eq:ex1a1}) and SBPs (\ref{eq:ex1a0a0}) mentioned above.

To that end, let $A_i\in\mathbb{R}^{d\times d}, i\in{1,2,\dots,n}$ be random matrices that are mutually independent, with the components of each $A_i$ being independent standard normals. We consider symmetric standard normal matrices $\bar{A}_i\in\mathbb{R}^{d\times d}, i\in{1,2,\dots,n}$ of the form
 \begin{eqnarray}
\bar{A}_i = \frac{1}{\sqrt{2}} \lp  A_i + A_i^T\rp.
\end{eqnarray}
As stated earlier, random matrix discrepancy problem is then the following 
\begin{eqnarray}
\hspace{-1.5in}\mbox{$\mathbf{\mathcal {MD}}(G,\alpha)$:} \hspace{1in} \kappa = \min_{\x} & & \left \| \sum_{i=1}^{n } \bar{A}_i\x_i \right \|_{2} \nonumber \\
\mbox{subject to}
 & & \x\in \left \{-\frac{1}{\sqrt{n}},\frac{1}{\sqrt{n}} \right \}^n, \label{eq:ex1a1a1}
\end{eqnarray}
 where $\|\cdot\|_2$ for symmetric matrix argument assumes its standard operator norm, i.e., $\| A\|_2 = \max_{\|\y\|_2=1} \y^TA\y$. One can then rewrite (\ref{eq:ex1a1a1}) as 
\begin{eqnarray}
\kappa =  
   \frac{1}{\sqrt{d}} \min_{\x\in \mB^n} \max_{\y\in\mS^d}   \y^T \lp \sum_{i=1}^{n} \bar{A}_i\x_i \rp  \y 
 =
   \sqrt{\frac{2}{d}}\, \xi \quad \mbox{with}\quad \xi\triangleq \min_{\x\in \mB^n} \max_{\y\in\mS^d}   \y^T \lp \sum_{i=1}^{n} A_i\x_i \rp  \y ,
 \label{eq:ex3}
\end{eqnarray}
where $\mS^m$ is the $m$-dimensional unit sphere  (i.e., $\mS^m=\{\y|\|\y\|_2=1\}$).

 The above optimization formulations can be embedded into a free energy context as well. For array $\cA = \{A_i\}_{i=1,\dots,n}$ consider   Hamiltonian
\begin{equation}
\cH(\cA)= \y^T \lp \sum_{i=1}^{n}A_i\x_i \rp \y,\label{eq:ham1}
\end{equation}
and associate to it the following partition function
\begin{equation}
Z(\beta,\cA)=\sum_{\x\in\cX} \lp \sum_{\y\in\cY}e^{\beta   \cH(\cA)  }\rp^{-1}.  \label{eq:partfun}
\end{equation}
For the time being the sets $\cX$ and $\cY$ are general; later on we will specialize them to fit the problems of interest here. One can then introduce  the thermodynamic limit of the corresponding (``\emph{reciprocal}'') average free energy
\begin{eqnarray}
f_{sq}(\beta) & = & \lim_{n\rightarrow\infty}\frac{\mE_\cA\log{(Z(\beta,\cA)})}{\beta  }
=\lim_{n\rightarrow\infty} \frac{\mE_\cA\log\lp \sum_{\x\in\cX} \lp \sum_{\y\in\cY}e^{\beta \lp   \cH(\cA)   \rp   }\rp^{-1}\rp}{\beta } \nonumber \\
& = &\lim_{n\rightarrow\infty} \frac{\mE_\cA\log\lp \sum_{\x\in\cX} \lp \sum_{\y\in\cY}e^{\beta \lp   \y^T\sum_{i=1}^{n} A_i\x_i \y   \rp  )}\rp^{-1}\rp}{\beta }.\label{eq:logpartfunsqrt}
\end{eqnarray}
Throughout the paper we adopt the convention that the subscripts next to $\mP$ and/or $\mE$ (if present) denote the source of underlying randomness (in (\ref{eq:logpartfunsqrt}),  it is $\cA$). 

Taking $\beta\rightarrow \infty$, we arrive at the ground state energy (GSE)
\begin{eqnarray}
f_{sq}(\infty)   \triangleq    \lim_{\beta\rightarrow\infty}f_{sq}(\beta) & = &
\lim_{\beta,n\rightarrow\infty}\frac{\mE_\cA\log{(Z(\beta,\cA)})}{\beta }
=
 \lim_{n\rightarrow\infty} \mE_\cA \max_{\x\in\cX}  -  \max_{\y\in\cY}  \y^T \lp \sum_{i=1}^{n} A_i\x_i \rp \y    \nonumber \\
& = & - \lim_{n\rightarrow\infty} \mE_\cA \min_{\x\in\cX}  \max_{\y\in\cY}  \y^T \lp \sum_{i=1}^{n}A_i\x_i \rp\y  = - \lim_{n\rightarrow\infty} \mE_\cA  \xi_0(\cX,\cY) ,
  \label{eq:limlogpartfunsqrta0}
\end{eqnarray}
with
\begin{eqnarray}
\xi_0(\cX,\cY) & \triangleq &  \min_{\x\in\cX}  \max_{\y\in\cY}  \y^T \lp\sum_{i=1}^{n} A_i\x_i\rp\y .
  \label{eq:limlogpartfunsqrta0b0}
\end{eqnarray}
Clearly, since  $\xi_0(\cX,\cY) $ and  $\xi$  are directly connected, so are $f_{sq}(\infty)$ and $\xi$. In other words, determining GSE  $f_{sq}(\infty)$ is sufficient to determine $\xi$.

\subsection{Information theoretic and algorithmic $\kappa$ -- computational gaps}
 \label{sec:caprole}

Of our key interest is the so-called margin $\kappa$ from (\ref{eq:ex1a1a1}) or alternatively $\xi$ from (\ref{eq:ex3}) (since all relevant quantities will concentrate, we often interchangeably refer to $\xi$ and $\mE\, \xi$). As emphasized on several occasions, we consider large dimensional proportional regime where
 \begin{eqnarray}\label{eq:ex4a0}
\alpha & = &    \lim_{n\rightarrow \infty} \frac{n}{d^2}  ,
\end{eqnarray}
and distinguish the following two values for $\kappa$
 \begin{align}
 \kappa_{inf} \triangleq & \, \kappa_{inf}(\alpha) 
 =  \max \{\kappa \,|\hspace{.08in}  \lim_{n\rightarrow\infty}\mP_\cA\lp \kappa \hspace{.07in}\mbox{ is achievable} \rp\longrightarrow 1\} 
\nonumber \\
 \kappa_{alg}  \triangleq &\,  \kappa_{alg}(\alpha) 
 =  \max \{\kappa \,|\hspace{.08in}  \lim_{n\rightarrow\infty}\mP_\cA\lp \kappa \hspace{.07in}\mbox{ is achievable in polynomial time} \rp\longrightarrow 1\} .  \label{eq:ex4}
\end{align}

All our analytical considerations will be in a statistical context. To facilitate writing, we often skip reemphasizing  that statements hold with high probability. 
 
Practically speaking, the above definitions introduce $\kappa_{inf}$ and  $\kappa_{alg}$ as two potentially different margins $\kappa$. The former represents the margin without computational restrictions, whereas the latter represents the margin achievable in polynomial time. $\kappa_{inf}$ is typically referred to as the information theoretical value, whereas $\kappa_{alg}$ is typically referred to as the algorithmic value. It is immediate that  
\begin{eqnarray}
\kappa_{alg}\leq \kappa_{inf}.
  \label{eq:ex4a1}
\end{eqnarray}
On the other hand, it is far from obvious whether the inequality is strict. If it is, one has nonzero \emph{statistical-computational gap} (SCG)
 \begin{eqnarray}
SCG \triangleq  \kappa_{inf} - \kappa_{alg}.
  \label{eq:ex4a2}
\end{eqnarray}
Settling the above question regarding inequality strictness is an extraordinary challenge with direct implications on classical and statistical P vs NP. We below present a plethora of results that shed a new light and further our understanding of this phenomenon and its particularities within the matrix discrepancy context.

\section{Characterization of $\kappa_{inf}$ via RDT}
\label{sec:implem}

We start by recall on the four key RDT principles \cite{StojnicCSetam09,StojnicICASSP10var,StojnicISIT2010binary,StojnicRegRndDlt10} (for more on additional upgrades and associated algorithmic and neural networks implications, see, e.g., \cite{Stojnicalgbp25,Stojnicclupsk25,Stojnicalgsbp26,StojnicNN27,Stojnictcmspnncaprdt23,Stojnictcmspnncapliftedrdt23}):

\begin{enumerate}

\item \emph{Finding underlying optimization algebraic representation}

\item \emph{Determining random dual} 

\item \emph{Handling random dual}

\item \emph{Double-checking strong random duality.}

\end{enumerate}
We discuss below all four principles and their connections to the problems of interest here.

\subsection{Finding fitting algebraic representation} 
\label{sec:randpr}

Keeping in mind  (\ref{eq:ex3}) we consider $m=d^2$ and set
\begin{eqnarray}\label{eq:amat0a1}
\cX & = & \left \{-\frac{1}{\sqrt{n}},\frac{1}{\sqrt{n}} \right \}^n =\mB^n
 \nonumber \\
\cY & = & \cY_{MD} \triangleq \{ Y \,| \, Y=Y^T\in\mR^{d\times d},\, \|Y\|_F=1, \,\mbox{rank}(Y)=1   \}
.
 \end{eqnarray}
Combining   (\ref{eq:ex3}) and  (\ref{eq:amat0a1}), we also have  
\begin{equation}\label{eq:algb1}
\xi = \min_{\x\in \mB^n} \max_{\y\in\cS^d}   \y^T \lp \sum_{i=1}^{n} A_i\x_i \rp  \y 
= \min_{\x\in \mB^n} \max_{Y\in\cY_{MD}}  \,  \tr \lp \lp \sum_{i=1}^{n} A_i\x_i \rp Y\rp .
 \end{equation}
Within the RDT,  the optimization in (\ref{eq:algb1}) is called \emph{random primal} .

\subsection{Determining random dual} 
\label{sec:randdual}

The following theorem complements the above random primal  with the so-called \emph{random dual}.

\begin{theorem}
\label{thm:thm1}
 Let elements of $H\in\mR^{d\times d}$ and $\g\in\mR^n$ be standard normal random variables independent among themselves and of any other randomness. Set
\begin{eqnarray}
   \label{eq:thm1eq1}
 L & = &  
\min_{\x\in\mB^n} \max_{\y\in\cY_{MD}}  \lp \tr(HY) +\g^T\x \rp  .
\end{eqnarray}
Let $\cA = \{A_i\}_{i=1,\dots,n}$ be an array of $n$ independent standard normal matrices (entries of each $A_i$ are independent as well). Let $\xi$ be as in (\ref{eq:algb1}).  One then has  
\begin{eqnarray}
   \label{eq:thm1eq2}
\lim_{n\rightarrow \infty}    \mE \xi    \leq  \lim_{n\rightarrow \infty}   \mE L  ,
\end{eqnarray}
with the righthand side being the random dual.
\end{theorem}

\begin{proof} Let $g$ be a standard normal random variable (independent of all other randomness).  Consider two centered Gaussian processes indexed by an array $\cX = \{\x,Y\}$
  \begin{eqnarray}
\label{eq:mr1}
 \cG (\cX) & \triangleq &  \cG (\x,Y)  \triangleq   \tr\lp \lp \sum_{i=1}^{m}A_i\x_i  \rp Y\rp +  g  \nonumber   \\
 \cG_l (\cX) & \triangleq &  \cG_l (\x,Y)  \triangleq    \tr(HY) +\g^T\x.
  \end{eqnarray}
 For two arrays $\cX^{(1)}=\{ \x^{(1)},Y^{(1)}\}$ and $\cX^{(2)}=\{ \x^{(2)},Y^{(2)}\}$ with $ \x^{(i)} \in \mB^n$ and $ Y^{(i)} \in \cY_{MD}$, $i=1,2$,  we first write
  \begin{eqnarray}
\label{eq:mr2}
\mE \cG (\cX^{(1)})\cG (\cX^{(2)})   & =  &   \tr \lp \lp Y^{(1)}\rp^TY^{(2)} \rp \lp\x^{(2)}\rp^T\x^{(1)} + 1
\nonumber   \\
\mE \cG_l (\cX^{(1)})\cG_l (\cX^{(2)})  & =  &   \tr \lp \lp Y^{(1)}\rp^TY^{(2)} \rp + \lp\x^{(2)}\rp^T\x^{(1)} .
  \end{eqnarray}
From (\ref{eq:mr2}), we further find
  \begin{align}
\label{eq:mr5}
  \mE & \cG (\cX^{(1)})\cG (\cX^{(2)})
 -
\mE \cG_l (\cX^{(1)})\cG_l (\cX^{(2)} ) =
\nonumber
\\
 & =       
 \tr \lp \lp Y^{(1)}\rp^TY^{(2)} \rp \lp\x^{(2)}\rp^T\x^{(1)} + 1
- \tr \lp \lp Y^{(1)}\rp^TY^{(2)} \rp - \lp\x^{(2)}\rp^T\x^{(1)}  \nonumber \\
  & =       
\lp 1 - \tr \lp \lp Y^{(1)}\rp^TY^{(2)} \rp \rp \lp 1 - \lp\x^{(2)}\rp^T\x^{(1)}\rp
 \nonumber \\
&  \geq   0.  
 \end{align}
One then also observes
  \begin{align}
\label{eq:mr5a0}
  \mE & \cG (\cX^{(1)})\cG (\cX^{(1)})
 -
\mE \cG_l (\cX^{(1)})\cG_l (\cX^{(1)} )
   = 
\lp 1 - \tr \lp \lp Y^{(1)}\rp^TY^{(1)} \rp \rp \lp 1 - \lp\x^{(1)}\rp^T\x^{(1)}\rp
  =   0,
  \end{align}
  where the last equality follows since  $\|\x^{(1)}\|^2_2 = \lp \x^{(1)}\rp^T \x^{(1)}= \tr\lp \lp Y^{(1)}\rp^T Y^{(1)}\rp=\|Y^{(1)}\|^2_F = 1$.

Recalling on Theorem 1.1 from \cite{Gordon85} allows to complete the proof.

\begin{theorem}(\cite{Gordon85})
\label{thm:Gordonpos1} Let $X_{ij}$ and $Y_{ij}$, $1\leq i\leq n$, $1\leq j\leq m$, be two centered Gaussian processes which satisfy the following inequalities for all choices of indices
\begin{enumerate}
\item $\mE(X_{ij}^2)=\mE(Y_{ij}^2)$
\item $\mE(X_{ij}X_{jk})\geq \mE(Y_{ij}Y_{ik})$
\item $\mE(X_{ij}X_{lk})\leq \mE(Y_{ij}Y_{lk}), i\neq l$.
\end{enumerate}
 Then
\begin{equation*}
\mE(\min_{i} \max_{j} X_{i})\leq \mE(\min_i \max_j Y_{ij})  .
\end{equation*}
\end{theorem}

We now note the correspondence $Y\leftrightarrow\cG$ and $X\leftrightarrow\cG_l$. Applying  Theorem \ref{thm:Gordonpos1} to processes $\cG(\cdot)$ and  $\cG_l(\cdot)$ while keeping in mind such a correspondence gives
\begin{align}\label{eq:mt5a1a0}
&  &\mE \min_{\x\in\mB^n}\max_{Y\in\cY_{MD}} \cG(\x,Y)  & \leq \mE  \min_{\x\in\mB^n}\max_{Y\in\cY_{MD}} \cG_l(\x,Y) 
\nonumber \\
\Longleftrightarrow & & \mE  \min_{\x\in\mB^n}\max_{Y\in\cY_{MD}} 
 \lp  \tr \lp \lp \sum_{i=1}^{n} A_i\x_i \rp Y\rp +  \g \rp
 & \leq \mE  \min_{\x\in\mB^n}\max_{Y\in\cY_{MD}}    \lp  \tr(HY) +\g^T\x \rp
\nonumber \\
\Longleftrightarrow & & \mE  \min_{\x\in\mB^n}\max_{Y\in\cY_{MD}}  
   \tr \lp \lp \sum_{i=1}^{n} A_i\x_i \rp Y\rp
 & \leq \mE  \min_{\x\in\mB^n}\max_{Y\in\cY_{MD}}   \lp  \tr(HY) +\g^T\x \rp
 .
 \end{align}
Connecting  (\ref{eq:algb1}) and (\ref{eq:mt5a1a0}), one obtains
\begin{eqnarray}\label{eq:mt5a1a1}
 \lim_{n\rightarrow \infty}    \mE \, \xi  & \leq &   \lim_{n\rightarrow \infty} \mE  \min_{\x\in\mB^n}\max_{Y\in\cY_{MD}}   \lp  \tr(HY) +\g^T\x\rp
 ,
 \end{eqnarray}
which, together with (\ref{eq:thm1eq1}), produces  (\ref{eq:thm1eq2}) and completes the proof.
\end{proof}

The above results could potentially be further improved through the utilization of the lifted methodology \cite{Stojnictcmspnncapliftedrdt23}. We have the following theorem.

\begin{theorem}
\label{thm:thm4}
Assume Theorem \ref{thm:thm1} setup. Consider large dimensional proportional setting with $\alpha=\lim_{n\rightarrow \infty} \frac{n}{d^2}$. For $c_3>0$ set
\begin{eqnarray}
   \label{eq:thm4eq1}
 L^{(2)} & = &  
 \max_{c_3>0}\lp 
  \frac{c_{3}}{2} 
        -
  \frac{1}{c_3}\log \lp
    \mE e^{ -c_{3}  \min_{\x\in\mB^n }   \lp  \g^T\x \rp    }
        \rp
        - \frac{1}{c_3}\log \lp
    \mE e^{ -c_{3}  \max_{\y\in\cY_{MD}}   \tr\lp HY  \rp    }
        \rp
\rp .
\end{eqnarray}
Then one has  
\begin{eqnarray}
   \label{eq:thm4eq2}
\lim_{n\rightarrow \infty}    \mE\, \sqrt{\frac{2}{d}}\,\xi    \geq  \lim_{n\rightarrow \infty}   \mE \,\sqrt{\frac{2}{d}}\, L^{(2)} \geq  \lim_{n\rightarrow \infty}   \mE \, \sqrt{\frac{2}{d}}\,L ,
\end{eqnarray}
with the middle term being the lifted random dual.
\end{theorem}

\begin{proof}
First, we also recall on Corollary 1.3 from \cite{Gordon85}.   
\begin{theorem}(\cite{Gordon85})
\label{thm:Gordoncor1} Assume the setup of Theorem \ref{thm:Gordonpos1}. Let $\psi_q(\cdot)$ be an increasing function on the real axis. Then
\begin{equation*}
E(\min_{i}\max_{j}\psi_q(X_{i}))\leq E(\min_i\max_{j} \psi_q(Y_{i})) .
\end{equation*}
Moreover,  let $\psi_q(\cdot)$ be a decreasing function on the real axis. Then
\begin{equation*}
E(\max_{i}\min_{j}\psi_q(X_{i}))\geq E(\max_i\min_{j} \psi_q(Y_{i})) .
\end{equation*}
\end{theorem}

For $\psi_q(x)=x$, one has that Theorem \ref{thm:Gordoncor1} matches Theorem \ref{thm:Gordonpos1}. The key question is then whether $\psi_q(\cdot)$ better than linear can be found. We discover that $\psi_q(x)=e^{-c_3 x}$, with $c_3>0$, is an excellent choice.

Recalling again the correspondence $Y\leftrightarrow\cG$ and $X\leftrightarrow\cG_l$ and combining (\ref{eq:mr5}) and (\ref{eq:mr5a0}) with  Theorem \ref{thm:Gordoncor1}   gives the following for $\psi_q(x)=e^{-c_3 x},c_3>0$
\begin{eqnarray}
\label{eq:mr6}
  \mE \max_{\x\in\mB^n} \min_{ Y\in\cY_{MD}} e^{-c_{3}\cG_l(\x,Y)} &\geq &    \mE \max_{\x\in\mB^n}\min_{ Y\in\cY_{MD}} e^{-c_{3}\cG(\x,Y)} .
\end{eqnarray}
From (\ref{eq:mr1}) and (\ref{eq:mr6}) one then finds
\begin{eqnarray}
\label{eq:mr7}
    \mE \max_{\x\in\mB^n} \min_{ Y\in\cY_{MD}}  e^{ - c_{3}  \lp \tr(HY) +\g^T\x \rp   } &\geq & 
        \mE \max_{\x\in\mB^n} \min_{ Y\in\cY_{MD}} 
      e^{ - c_{3} \lp \tr\lp \lp \sum_{i=1}^{m}A_i\x_i  \rp Y\rp +  g \rp   },
  \end{eqnarray}
which is equivalent to
\begin{eqnarray}
\label{eq:mr8}
    \mE \max_{\x\in\mB^n} \min_{ Y\in\cY_{MD}}  e^{ - c_{3}  \lp \tr(HY) +\g^T\x \rp   } &\geq & 
   e^{\frac{c_{3}^2}{2}}
        \mE \max_{\x\in\mB^n} \min_{ Y\in\cY_{MD}}  
        e^{ - c_{3} \lp \tr\lp \lp \sum_{i=1}^{m}A_i\x_i  \rp Y\rp \rp   },
\end{eqnarray}
and
\begin{eqnarray}
\label{eq:mr9}
  \log \lp  \mE \max_{\x\in\mB^n} \min_{ Y\in\cY_{MD}}  e^{ - c_{3}  \lp \tr(HY) +\g^T\x \rp   } 
  \rp
  & \geq & 
   \frac{c_{3}^2}{2} +
   \log \lp
        \mE \max_{\x\in\mB^n} \min_{ Y\in\cY_{MD}}  
        e^{ - c_{3} \lp \tr\lp \lp \sum_{i=1}^{m}A_i\x_i  \rp Y\rp \rp   }  
        \rp .
\end{eqnarray}
Applying Jensen's inequality to the right-hand side and dividing both sides by $c_3$ gives
\begin{eqnarray}
\label{eq:mr9a0}
 \frac{1}{c_3} \log \lp  \mE \max_{\x\in\mB^n} \min_{ Y\in\cY_{MD}}  e^{ - c_{3}  \lp \tr(HY) +\g^T\x \rp   } 
  \rp
  & \geq & 
   \frac{c_{3}}{2} +
          \mE \max_{\x\in\mB^n} \min_{ Y\in\cY_{MD}}  
        -  \tr\lp \lp \sum_{i=1}^{m}A_i\x_i  \rp Y\rp  .
\end{eqnarray}
One then trivially has
\begin{eqnarray}
\label{eq:mr9a1}
 \frac{1}{c_3} \log \lp  \mE \max_{\x\in\mB^n} \min_{ Y\in\cY_{MD}}  e^{ - c_{3}  \lp \tr(HY) +\g^T\x \rp   } 
  \rp
  & \geq & 
   \frac{c_{3}}{2} -
          \mE \min_{\x\in\mB^n} \max_{ Y\in\cY_{MD}}  
         \tr\lp \lp \sum_{i=1}^{m}A_i\x_i  \rp Y\rp  ,
\end{eqnarray}
and after utilizing  independence of $H$ and $\g$
\begin{eqnarray}
\label{eq:mr9a2}
 \frac{1}{c_3} \log \lp  \mE \max_{\x\in\mB^n}  e^{ - c_{3}   \g^T\x    } 
  \rp
+
 \frac{1}{c_3} \log \lp  \mE \min_{ Y\in\cY_{MD}}  e^{ - c_{3}  \lp \tr(HY)  \rp   } 
  \rp
  & \geq & 
   \frac{c_{3}}{2} -
          \mE \min_{\x\in\mB^n} \max_{ Y\in\cY_{MD}}  
         \tr\lp \lp \sum_{i=1}^{m}A_i\x_i  \rp Y\rp  . \nonumber \\
\end{eqnarray}
A simple rearrangement then also gives
\begin{align}
\label{eq:mr9a3}
          \mE \min_{\x\in\mB^n} \max_{ Y\in\cY_{MD}}  
         \tr\lp \lp \sum_{i=1}^{m}A_i\x_i  \rp Y\rp 
  & \geq 
   \frac{c_{3}}{2} -
  \frac{1}{c_3} \log \lp  \mE \max_{\x\in\mB^n}  e^{ - c_{3}   \g^T\x    } 
  \rp
-
 \frac{1}{c_3} \log \lp  \mE \min_{ Y\in\cY_{MD}}  e^{ - c_{3}   \tr(HY)    } 
  \rp  \nonumber \\
  & =  
   \frac{c_{3}}{2} -
  \frac{1}{c_3} \log \lp  \mE   e^{ - c_{3} \min_{\x\in\mB^n}  \g^T\x    } 
  \rp
-
 \frac{1}{c_3} \log \lp  \mE  e^{ - c_{3}  \max_{ Y\in\cY_{MD}}  \tr(HY)     } 
  \rp.
\end{align}
Since (\ref{eq:mr9a3}) holds for any $c_3>0$, it holds for the ``worst-case'' one as well. In other words, one also has
\begin{eqnarray}
\label{eq:mr10}
\mE\,\xi  & = & \min_{\x\in \mB^n} \max_{Y\in\cY_{MD}}  \,  \tr \lp \lp \sum_{i=1}^{n} A_i\x_i \rp Y\rp 
\nonumber \\
& \geq &\max_{c_3>0}
   \lp
   \frac{c_{3}}{2} -
  \frac{1}{c_3} \log \lp  \mE   e^{ - c_{3} \min_{\x\in\mB^n}  \g^T\x    } 
  \rp
-
 \frac{1}{c_3} \log \lp  \mE  e^{ - c_{3}  \max_{ Y\in\cY_{MD}} \tr(HY)  } 
  \rp
  \rp,
\end{eqnarray}
which matches the first inequality in (\ref{eq:thm4eq1}). One then observes that the second inequality in (\ref{eq:thm4eq1}) follows automatically for $c_3\rightarrow 0$ which completes the theorem's proof.
\end{proof}

\subsection{Handling lifted random dual}
\label{sec:hlrd}

We start by observing
\begin{eqnarray}
   \label{eq:hlrd1}
\min_{\x\in\mB^n }  \,   \g^T\x    & = &   
 -\frac{1}{\sqrt{n}} \sum_{i=1}^{n} |\g_i|
  \nonumber \\
 \max_{\y\in\cY_{MD}}   \tr \lp  HY  \rp
  & = &   \| H \|_2.
\end{eqnarray}
One then finds
\begin{eqnarray}
   \label{eq:hlrd3}
\frac{1}{c_3}\log \lp
    \mE e^{ -c_{3}  \min_{\x\in\mB^n }   \g^T\x      }
        \rp
&  = &
\frac{1}{c_3}\log \lp
    \mE e^{  \frac{c_{3}}{\sqrt{n}} \sum_{i=1}^{n} |\g_i|    }
        \rp
        =
        \frac{n}{c_3}\log \lp
    \mE e^{ \frac{c_{3}}{\sqrt{n}} |\g_1|    }
        \rp
        \nonumber \\
        & = & 
     \frac{n}{c_3}\log 
     \lp
       {\mathrm{e}}^{\frac{{c_{3}}^2}{2\,n}}\, \mathrm{erfc}\left(-\frac{ c_{3}}{\sqrt{2n}}\right) 
      \rp   
= \frac{c_3}{2} +
     \frac{n}{c_3}\log 
     \lp
        \mathrm{erfc}\left(-\frac{ c_{3}}{\sqrt{2n}}\right)
      \rp  . \nonumber \\ 
 \end{eqnarray}
Moreover,
\begin{eqnarray}
   \label{eq:hlrd2a0}
\frac{1}{c_3}\log \lp
    \mE e^{ -c_{3}  \max_{Y\in\cY_{MD} }   \tr(HY)      }
        \rp
&  = &
\frac{1}{c_3}\log \lp
    \mE e^{ - \frac{c_{3}}{2}  \max_{Y\in\cY_{MD} }   \tr\lp\lp H+H^T\rp Y \rp      }
        \rp
\nonumber \\
&  = &
\frac{1}{c_3}\log \lp
    \mE e^{ - \frac{c_{3}\sqrt{d}}{\sqrt{2}}  \max_{Y\in\cY_{MD} }  \frac{1}{\sqrt{2d}} \tr\lp\lp H+H^T\rp Y \rp      }
        \rp
\nonumber \\
&  = &
\frac{1}{c_3}\log \lp
    \mE e^{ - \frac{c_{3}\sqrt{d}}{\sqrt{2}}  \| \bar{H}  \|_2     }
        \rp,
 \end{eqnarray}
where $\bar{H}= \frac{1}{\sqrt{2d}} \lp H+H^T\rp$. We set $h=\|\bar{H}\|_2$ and adopt scaling $c_3\rightarrow c_3^{(s)}d^{\frac{3}{2}}$ where $c_3^{(s)}$ does not change as $d$ and $n$ grow. Let $f_h$ be the pdf of $h$. For $h<2$, Maillard determined 
\begin{eqnarray}
   \label{eq:hlrd2a1}
\lim_{d\rightarrow\infty } \frac{1}{d^2}\log \lp f_{h}\rp   
=
\lp\frac{\ln\left(\frac{h}{2}\right)}{2}
-\frac{h^2}{8}+\frac{h^4}{128}+\frac{3}{8} \rp. 
\end{eqnarray} 
One then finds
\begin{eqnarray}
   \label{eq:hlrd2a2}
     \mE e^{ - \frac{c_{3}\sqrt{d}}{\sqrt{2}}  h     }
        =
     \mE e^{ - \frac{c_{3}^{)s)}d^2}{\sqrt{2}}  h     }
        =
        \int_{h} e^{ - \frac{c_{3}^{(s)}d^2}{\sqrt{2}}  h } f_h  d h         =
        \int_{h} e^{ - \frac{c_{3}^{(s)}d^2}{\sqrt{2}}  h } e^{d^2\,\frac{1}{d^2}\log\lp f_h \rp}  d h =
        \int_{h} e^{ d^2 \lp - \frac{c_{3}^{(s)}}{\sqrt{2}}  h + \frac{1}{d^2}\log\lp f_h \rp \rp }  d h .
 \end{eqnarray}
Utilizing the saddle point method and (\ref{eq:hlrd2a1}), we further find
\begin{eqnarray}
   \label{eq:hlrd2}
   \frac{1}{d^2} \log\lp \mE e^{ - \frac{c_{3}\sqrt{d}}{\sqrt{2}}  h  } \rp 
 \longrightarrow 
  \max_{h<2}\lp - \frac{c_{3}^{(s)}}{\sqrt{2}}  h + \frac{1}{d^2}\log\lp f_h \rp \rp .
 \end{eqnarray}  
A combination of (\ref{eq:thm4eq1}), (\ref{eq:hlrd3}), and (\ref{eq:hlrd2}) gives
 \begin{eqnarray}
\label{eq:hlrd4}
  L^{(2)} & = &  
 \max_{c_3>0}\lp 
  \frac{c_{3}}{2} 
        -
  \frac{1}{c_3}\log \lp
    \mE e^{ -c_{3}  \min_{\x\in\mB^n }   \lp  \g^T\x \rp    }
        \rp
        - \frac{1}{c_3}\log \lp
    \mE e^{ -c_{3}  \max_{\y\in\cY_{MD}}   \tr\lp HY  \rp    }
        \rp
\rp 
\nonumber \\
& = & 
\max_{c_3>0}\lp 
  \frac{c_{3}}{2} 
        -
\frac{c_3}{2} -
     \frac{n}{c_3}\log 
     \lp
        \mathrm{erfc}\left(-\frac{ c_{3}}{\sqrt{2n}}\right) 
      \rp
              - \frac{d^2}{c_3}
                \max_{h<2}\lp - \frac{c_{3}^{(s)}}{\sqrt{2}}  h + \frac{1}{d^2}\log\lp f_h \rp \rp 
\rp 
\nonumber \\
& = & 
\max_{c_3>0}\lp 
  -
     \frac{n}{c_3}\log 
     \lp
        \mathrm{erfc}\left(-\frac{ c_{3}}{\sqrt{2n}}\right) 
      \rp
              - \frac{d^2}{c_3}
                \max_{h<2}\lp - \frac{c_{3}^{(s)}}{\sqrt{2}}  h + \frac{1}{d^2}\log\lp f_h \rp \rp 
\rp 
.
\end{eqnarray}
Recalling $\alpha=\lim_{n\rightarrow \infty} \frac{n}{d^2}$, we further write
 \begin{eqnarray}
\label{eq:hlrd5}
 \lim_{d,n\rightarrow\infty}\sqrt{\frac{2}{d}} L^{(2)} 
 & = &  
\sqrt{2} \lim_{d,n\rightarrow\infty}\max_{c_3>0}\lp 
  -
     \frac{\alpha}{c_3^{(s)}}\log 
     \lp
        \mathrm{erfc}\left(-\frac{ c_{3}^{(s)} d^{\frac{3}{2}} }{\sqrt{2\alpha d^2}}\right) 
      \rp
              -  \frac{1}{c_3^{(s)}}
                \max_{h<2}\lp - \frac{c_{3}^{(s)}}{\sqrt{2}}  h + \frac{1}{d^2}\log\lp f_h \rp \rp 
\rp 
\nonumber \\
 & = &  
\sqrt{2} \lim_{d,n\rightarrow\infty}\max_{c_3>0}\lp 
  -
     \frac{\alpha}{c_3^{(s)}}\log 
     \lp
     2
      \rp
              -  \frac{1}{c_3^{(s)}}
                \max_{h<2}\lp - \frac{c_{3}^{(s)}}{\sqrt{2}}  h + \frac{1}{d^2}\log\lp f_h \rp \rp 
\rp 
\nonumber \\
 & = &  
\sqrt{2}  \max_{c_3>0}\lp 
  -
     \frac{\alpha}{c_3^{(s)}}\log 
     \lp
     2
      \rp
              -  \frac{1}{c_3^{(s)}}
                \max_{h<2}\lp - \frac{c_{3}^{(s)}}{\sqrt{2}}  h +
                \lp\frac{\ln\left(\frac{h}{2}\right)}{2}
-\frac{h^2}{8}+\frac{h^4}{128}+\frac{3}{8} \rp
 \rp 
\rp 
.  
\end{eqnarray}
   
The following theorem summarizes the preceding discussion.

\begin{theorem}
\label{thm:thm6}
Assume the setup of Theorem \ref{thm:thm4} with   $L^{(2)}$ as in
(\ref{eq:thm4eq1}). Then 
 \begin{equation}
   \label{eq:thm6eq1}
 \lim_{d,n\rightarrow\infty}\sqrt{\frac{2}{d}} L^{(2)} 
   =   
\sqrt{2}  \max_{c_3>0}\lp 
  -
     \frac{\alpha}{c_3^{(s)}}\log 
     \lp
     2
      \rp
              -  \frac{1}{c_3^{(s)}}
                \max_{h<2}\lp - \frac{c_{3}^{(s)}}{\sqrt{2}}  h +
                \lp\frac{\ln\left(\frac{h}{2}\right)}{2}
-\frac{h^2}{8}+\frac{h^4}{128}+\frac{3}{8} \rp
 \rp 
\rp 
,  
\end{equation}
and
\begin{eqnarray}
   \label{eq:thm6eq2}
\lim_{n\rightarrow \infty}    \mE\, \sqrt{\frac{2}{d}}\,\xi    \geq  \lim_{n\rightarrow \infty}   \mE \,\sqrt{\frac{2}{d}}\, L^{(2)} \geq  \lim_{n\rightarrow \infty}   \mE \, \sqrt{\frac{2}{d}}\,L 
.
\end{eqnarray}
\end{theorem}

\begin{proof}
Follows from the preceding discussion.
\end{proof} 

Theorem \ref{thm:thm6}
 is sufficient to numerically evaluate a lower bound on the minimum matrix discrepancy $\lim_{d,n\rightarrow \infty} \sqrt{\frac{2}{\pi}}, \mE \xi$. However one can be even a bit more explicit regarding analytical characterizations. Namely, let
 \begin{eqnarray}
 \label{eq:hrdexp1}
f_L( c_3^{(s)},h;\alpha ) = -
     \frac{\alpha}{c_3^{(s)}}\log 
     \lp
     2
      \rp
              -  \frac{1}{c_3^{(s)}}
                 \lp - \frac{c_{3}^{(s)}}{\sqrt{2}}  h +
                \lp\frac{\ln\left(\frac{h}{2}\right)}{2}
-\frac{h^2}{8}+\frac{h^4}{128}+\frac{3}{8} \rp
 \rp.
 \end{eqnarray}
 After taking the $c_3^{(s)}$ and $h$ derivatives we find
 \begin{eqnarray}
 \label{eq:hrdexp2}
\frac{d\,f_L( c_3^{(s)},h;\alpha ) }{d\,c^{(s)}} 
& = & 
     \frac{\alpha}{ \lp c_3^{(s)} \rp^2 }\log 
     \lp
     2
      \rp
              +  \frac{1}{ \lp c_3^{(s)} \rp^2 }
                \lp\frac{\ln\left(\frac{h}{2}\right)}{2}
-\frac{h^2}{8}+\frac{h^4}{128}+\frac{3}{8} \rp 
\nonumber \\
\frac{d\,f_L( c_3^{(s)},h;\alpha )}{d\, h} 
& = & 
               -  \frac{1}{c_3^{(s)}}
                 \lp - \frac{c_{3}^{(s)}}{\sqrt{2}}   +
                \lp  \frac{1  }{2h}
-\frac{h}{4}+\frac{h^3}{32} \rp
 \rp .
  \end{eqnarray}
Let $\bar{h}$ and $\bar{c}_3^{(s)}$ be the solutions of (\ref{eq:thm6eq1}). Then from (\ref{eq:hrdexp3}), we have
 \begin{eqnarray}
 \label{eq:hrdexp3}
\alpha \log \lp
     2
      \rp
              +                   \lp\frac{\ln\left(\frac{\bar{h}}{2}\right)}{2}
-\frac{\bar{h}^2}{8}+\frac{\bar{h}^4}{128}+\frac{3}{8} \rp & = & 0
\nonumber \\
 \bar{c}_{3}^{(s)} -  \sqrt{2}  
                \lp  \frac{1  }{2\bar{h}}
-\frac{\bar{h}}{4}+\frac{\bar{h}^3}{32} \rp
  & = & 0.
  \end{eqnarray}
Moreover, combining (\ref{eq:hrdexp2}) and  (\ref{eq:hrdexp3}), we find
 \begin{eqnarray}
 \label{eq:hrdexp4}
f_L( \bar{c}_3^{(s)},\bar{h};\alpha ) 
& = & -
     \frac{\alpha}{\bar{c}_3^{(s)}}\log 
     \lp
     2
      \rp
              -  \frac{1}{\bar{c}_3^{(s)}}
                 \lp - \frac{\bar{c}_{3}^{(s)}}{\sqrt{2}}  \bar{h} +
                \lp\frac{\ln\left(\frac{\bar{h}}{2}\right)}{2}
-\frac{\bar{h}^2}{8}+\frac{\bar{h}^4}{128}+\frac{3}{8} \rp
 \rp
 \nonumber \\
 & = & -
     \frac{\alpha}{\bar{c}_3^{(s)}}\log 
     \lp
     2
      \rp
              -  \frac{1}{\bar{c}_3^{(s)}}
                 \lp - \frac{\bar{c}_{3}^{(s)}}{\sqrt{2}}  \bar{h} 
                -\alpha\log(2)
 \rp
  \nonumber \\
 & = & 
 \frac{\bar{h}}{\sqrt{2}}
 .
 \end{eqnarray}

The above allows us to formulate the following corollary.

\begin{corollary}
\label{cor:cor1}
Assume the setup of Theorem \ref{thm:thm4}. Let $\bar{h}$ satisfy
 \begin{eqnarray}
 \label{eq:cor1eq1}
\alpha \log \lp
     2
      \rp
              +                   \lp\frac{\ln\left(\frac{\bar{h}}{2}\right)}{2}
-\frac{\bar{h}^2}{8}+\frac{\bar{h}^4}{128}+\frac{3}{8} \rp & = & 0
\end{eqnarray}
Then
\begin{eqnarray}
   \label{eq:cor1eq2}
\lim_{n\rightarrow \infty}    \mE\, \sqrt{\frac{2}{d}}\,\xi    \geq  \bar{h} \geq  \lim_{n\rightarrow \infty}   \mE \, \sqrt{\frac{2}{d}}\,L 
.
\end{eqnarray}
Also,
\begin{eqnarray}
   \label{eq:cor1eq3}
 \bar{c}_{3}^{(s)} =  \sqrt{2}  
                \lp  \frac{1  }{2\bar{h}}
-\frac{\bar{h}}{4}+\frac{\bar{h}^3}{32} \rp .
  \end{eqnarray}
\end{corollary}

\begin{proof}
Follows from the above discussion after one notes
 \begin{equation}
   \label{eq:cor1preq1}
 \lim_{d,n\rightarrow\infty}\sqrt{\frac{2}{d}} L^{(2)} =\sqrt{2}f_L( \bar{c}_3^{(s)},\bar{h};\alpha ) 
=\bar{h}.
\end{equation}
\end{proof} 
 
It is not that difficult to see that equation (\ref{eq:cor1eq1}) precisely matches Maillard's condition for lower-bounding $\kappa_{inf}$. This implies that our $\bar{h}$ is exactly Maillard's  estimate. However, as (\ref{eq:cor1eq2}) shows, precisely such $\bar{h}$ also lower-bounds $\lim_{n\rightarrow \infty}    \mE \, \kappa_{inf} = \lim_{n\rightarrow \infty}    \mE\, \sqrt{\frac{2}{d}}\,\xi $ which effectively means that not only does it match the Maillard's result, but it also does so as a lower bound on the true $\lim_{n\rightarrow \infty }\mE \, \kappa_{inf}$.

Due to highly non-convex nature of matrix discrepancy, the strong  random duality is clearly not in place.  To compensate for that one could implement the fully lifted RDT. However, conducting the associated numerical evaluations is fairly challenging. In the following sections, we instead opt for consideration of a proxy model that is more amenable to underlying numerics.

\section{Negative Hopfield model as a proxy for matrix discrepancy}
\label{sec:proxy1}

Recalling (\ref{eq:algb1}) we have
\begin{eqnarray}\label{eq:prx1}
 \xi & = &   \min_{\x\in \mB^n} \max_{Y\in\cY_{MD}}  \,  \tr \lp \lp \sum_{i=1}^{n} A_i\x_i \rp Y\rp 
= \min_{\x\in \mB^n} \max_{Y\in\cY_{MD}}  \,  \lp \sum_{i=1}^{n} \tr(A_iY)\x_i \rp 
\nonumber \\
& = & \frac{1}{\sqrt{2}}\min_{\x\in \mB^n} \max_{Y\in\cY_{MD}}  \,  \lp \sum_{i=1}^{n} \tr\lp \frac{1}{\sqrt{2}}(A_i+A_i^T)Y\rp\x_i \rp 
=
\frac{1}{\sqrt{2}}\min_{\x\in \mB^n} \max_{Y\in\cY_{MD}}  \,  \lp \sum_{i=1}^{n} \tr\lp \bar{A}_iY\rp\x_i \rp .
 \end{eqnarray}
Since ew consider large dimensional $d\rightarrow \infty $ settings, diagonal elements play no role. One can then drop $\mathrm{rank}(Y)=1$ constraint to have the following proxy for the traces
\begin{eqnarray}\label{eq:prx2}
 \tr(\bar{A}_iY) \longleftrightarrow 2\lp\a^{(i)}\rp^T\y ,
 \end{eqnarray}
where $\a^{(i)}\in\mR^{\frac{d(d-1)}{2}}$ are (independent) vectors of independent standard normals and $\y\in\mS^{\frac{d(d-1)}{2}}$. Keeping in mind that  $d\rightarrow \infty $ implies $\frac{d(d-1)}{2}/\frac{d^2}{2}\rightarrow  1$, we then have the following $\xi_-$ proxy in place of 
\begin{eqnarray}\label{eq:prx3}
 \xi_- & = &  a \sqrt{2} \min_{\x\in \mB^n} \max_{\y\in\mS^{\frac{d^2}{2}}}  \,  \lp \sum_{i=1}^{n}  \y^T \a^{(i)}\x_i \rp 
 =
 a \sqrt{2n} \frac{1}{\sqrt{n}}\min_{\x\in \mB^n} \max_{\y\in\mS^{\frac{d^2}{2}}}  \,  \y^TA\x,
 \end{eqnarray}
where $A\in\mR^{\frac{d^2}{2}\times n}$ is comprised of independent standard normals and $a$ is a scaling factor. Since we are interested in proportional regime where $d^2\sim n$ one has 
\begin{eqnarray}\label{eq:prx3a0}
 \frac{1}{\sqrt{n}}\min_{\x\in \mB^n} \max_{\y\in\mS^{\frac{d^2}{2}}}  \,  \y^TA\x\sim const. \quad \implies \quad a\sim \frac{1}{\sqrt{d}},
\end{eqnarray}
so that $\xi\sim \sqrt{d}$. This determines the scaling order of $a$. To get the corresponding constant $a_s$ such that $a=\frac{a_s}{\sqrt{d}}$, we further impose that the proxy should match the true value for $\alpha\rightarrow 0$ 
\begin{eqnarray}\label{eq:prx4}
\lim_{\alpha\rightarrow 0}\lim_{d,n\rightarrow \infty} \sqrt{\frac{2}{d}}\xi_- = 2.
 \end{eqnarray}
From (\ref{eq:prx3})  we first find 
\begin{eqnarray}\label{eq:prx5}
\lim_{\alpha\rightarrow 0}\lim_{d,n\rightarrow \infty} \sqrt{\frac{2}{d}}\xi_- 
=
\lim_{\alpha\rightarrow 0}\lim_{d,n\rightarrow \infty} \sqrt{\frac{2}{d}}
  a \sqrt{2n} \frac{1}{\sqrt{n}}\min_{\x\in \mB^n} \max_{\y\in\mS^{\frac{d^2}{2}}}  \,  \y^TA\x
  =
\lim_{\alpha\rightarrow 0}\lim_{d,n\rightarrow \infty} \sqrt{\frac{2}{d}}
  \frac{a_s}{\sqrt{d}} \sqrt{2n} \sqrt{\frac{d^2}{2n}} =a_s\sqrt{2} ,
 \end{eqnarray}
and then combining  (\ref{eq:prx4}) and (\ref{eq:prx5}), we obtain
\begin{eqnarray}\label{eq:prx6}
 a_s = \sqrt{2} .
 \end{eqnarray}
From (\ref{eq:prx3}) and (\ref{eq:prx6})
\begin{eqnarray}\label{eq:prx7}
\kappa_{inf,-} \triangleq \sqrt{\frac{2}{d}} \xi_-  =  
 2 \sqrt{2\alpha} \frac{1}{\sqrt{n}}\min_{\x\in \mB^n} \max_{\y\in\mS^m}  \,  \y^TA\x
 = 2 \sqrt{2\alpha} \xi_h \quad \mbox{ where } \quad \xi_h\triangleq \frac{1}{\sqrt{n}}\min_{\x\in \mB^n} \max_{\y\in\mS^m}  \,  \y^TA\x,
 \end{eqnarray}
with 
\begin{eqnarray}\label{eq:prx8}
m=\frac{d^2}{2}, \quad \quad \alpha_p\triangleq \frac{m}{n}=\frac{1}{2\alpha}, 
\end{eqnarray}
and $A\in\mR^{m\times n}$ comprised of independent standard normal elements. Modulo $2\sqrt{2\alpha} $ term, the above is classical negative Hopfield model (square root variant) with loading $\alpha_p=\frac{1}{2\alpha}$. Even though analysis of such a model is still quite challenging, as we will see below, its associated fully lifted RDT implementation is significantly more manageable. We also note for completeness that it can be shown via trivial Frobenius-operator norm inequality that  $\frac{1}{2}\kappa_{inf,-}\leq \kappa_{inf}$. We skip discussing details of these trivialities as our predominant interest in using the above proxy is its association with the algorithmic upper-bounding.

\section{Characterizing $\xi_h$ via RDT}
\label{sec:implemh}

We start with several RDT technical preliminaries. 

\subsection{Free energy interpretation}
\label{sec:freng}

Similarly to what we have done for matrix discrepancy in (\ref{eq:ham1}) -- (\ref{eq:limlogpartfunsqrta0b0}), we provide a corresponding free energy interpretation for negative Hopfield model. We parallel formulations presented in earlier section but proceed in a faster fashion by avoiding unnecessary repetitions whenever possible. 

As in (\ref{eq:ham1}), we first define negative Hopfield model Hamiltonian
\begin{equation}
\cH(A)= \y^T A\x,\label{eq:frham1}
\end{equation}
and associated partition function
\begin{equation}
Z(\beta,A)=\sum_{\x\in\mB^n} \lp \sum_{\y\in\mS^m}e^{\beta   \cH(A)  }\rp^{-1}.  \label{eq:frpartfun}
\end{equation}
We then have for rhe thermodynamic limit of the (``\emph{reciprocal}'') average free energy 
\begin{eqnarray}
f_{sq,h}(\beta) & = & \lim_{n\rightarrow\infty}\frac{\mE_A\log{(Z(\beta,A)})}{\beta\sqrt{n}  }
=\lim_{n\rightarrow\infty} \frac{\mE_A\log\lp \sum_{\x\in\mB^n} \lp \sum_{\y\in\mS^m}e^{\beta \lp   \cH(A)   \rp   }\rp^{-1}\rp}{\beta\sqrt{n} } \nonumber \\
& = &\lim_{n\rightarrow\infty} \frac{\mE_A\log\lp \sum_{\x\in\mB^n} \lp \sum_{\y\in\mS^m}e^{\beta \y^TA\x )}\rp^{-1}\rp}{\beta\sqrt{n} }.\label{eq:frlogpartfunsqrt}
\end{eqnarray}
After taking $\beta\rightarrow \infty$, we arrive at the ground state energy (GSE) (basically a negative Hopfield model analogue to (\ref{eq:frlimlogpartfunsqrta0}))
\begin{eqnarray}
f_{sq,h}(\infty)   \triangleq    \lim_{\beta\rightarrow\infty}f_{sq,h}(\beta) & = &
  - \lim_{n\rightarrow\infty} \mE_A \frac{1}{\sqrt{n}} \min_{\x\in\mB^n}  \max_{\y\in\mS^m}  \y^TA\x = -\xi_h.
  \label{eq:frlimlogpartfunsqrta0}
\end{eqnarray}
Clearly, determining GSE  $f_{sq,h}(\infty)$ is sufficient to determine $\xi_h$. However,  computing $f_{sq,h}(\infty)$ is typically not easy. One usually first studies $f_{sq,h}(\beta)$ for a general $\beta$ and then eventually specializes to $\beta\rightarrow\infty$ GSE regime. Keeping in mind that this specialization will be applied, throughout the analyses, we may on occasion neglect some of the GSE irrelevant terms.

\subsection{RDT basics}
\label{sec:randlincons}

Since the free energy given in (\ref{eq:frlogpartfunsqrt}),
\begin{eqnarray}
f_{sq,h}(\beta) & = &\lim_{n\rightarrow\infty} \frac{\mE_A\log\lp \sum_{\x\in\mB^n} \lp \sum_{\y\in\mS^m}e^{\beta  \y^TA\x }\rp^{-1}\rp}{\beta \sqrt{n}},\label{eq:hmsfl1}
\end{eqnarray}
is a statistics of random process $\y^TA\x$, a powerful mathematical machinery developed in \cite{Stojnicflrdt23} for studying such processes can be utilized. For practical implementation, we follow into the footsteps of \cite{Stojnicalgbp25,Stojnicalgsbp26}.  

First, we take a positive integer $r\in\mN$ and consider ordered vectors $\p=[\p_0,\p_1,\dots,\p_{r+1}]$, $\q=[\q_0,\q_1,\dots,\q_{r+1}]$, and $\c=[\c_0,\c_1,\dots,\c_{r+1}]$ with
 \begin{eqnarray}\label{eq:hmsfl2}
1=\p_0\geq \p_1\geq \p_2\geq \dots \geq \p_r\geq \p_{r+1} & = & 0 \nonumber \\
1=\q_0\geq \q_1\geq \q_2\geq \dots \geq \q_r\geq \q_{r+1} & = &  0,
 \end{eqnarray}
$\c_0=1$, $\c_{r+1}=0$. Let sets $\cX$ and $\cY$ be subsets of $\mS^n$  and  $\mS^m$, respectively (as earlier, $\mS^p$ stands for the unit sphere in $\mR^p$). For $k\in\{1,2,\dots,r+1\}$ we then set ${\mathcal U}_k\triangleq [u^{(4,k)},\u^{(2,k)},\h^{(k)}]$  where elements of  $u^{(4,k)}\in\mR$, $\u^{(2,k)}\in\mR^m$, and $\h^{(k)}\in\mR^n$ are standard normals independent among themselves and of all other randomness. For a given function $f_S(\cdot):\mR^n\rightarrow R$ and  $s\in\{-1,1\}$, we also set
  \begin{eqnarray}\label{eq:fl4}
\bar{\psi}_{S,\infty}(f_{S},\cX,\cY,\p,\q,\c,s)  =
 \mE_{G,{\mathcal U}_{r+1}} \frac{1}{n\c_r} \log
\lp \mE_{{\mathcal U}_{r}} \lp \dots \lp \mE_{{\mathcal U}_3}\lp\lp\mE_{{\mathcal U}_2} \lp \lp Z_{S,\infty}\rp^{\c_2}\rp\rp^{\frac{\c_3}{\c_2}}\rp\rp^{\frac{\c_4}{\c_3}} \dots \rp^{\frac{\c_{r}}{\c_{r-1}}}\rp, \nonumber \\
 \end{eqnarray}
with
\begin{eqnarray}\label{eq:fl5}
Z_{S,\infty} & \triangleq & e^{D_{0,S,\infty}} \nonumber \\
 D_{0,S,\infty} & \triangleq  & \max_{\x\in\cX } s \max_{\y\in\cY }
 \lp \sqrt{n} f_{S}
+\sqrt{n}      \lp\sum_{k=2}^{r+1}c_k\h^{(k)}\rp^T\x
+ \sqrt{n}  \y^T\lp\sum_{k=2}^{r+1}b_k\u^{(2,k)}\rp \rp \nonumber  \\
 b_k & \triangleq & b_k(\p,\q)=\sqrt{\p_{k-1}-\p_k} \nonumber \\
c_k & \triangleq & c_k(\p,\q)=\sqrt{\q_{k-1}-\q_k}.
 \end{eqnarray}
It is worth noting that $\bar{\psi}_{S,\infty}(f_{S},\cX,\cY,\p,\q,\c,s) $ form in (\ref{eq:fl4}) is practically useful for $r\geq 2$. For notational convenience, we say that when $r=1$ $\bar{\psi}_{S,\infty}(f_{S},\cX,\cY,\p,\q,\c,s) $ is obtained from (\ref{eq:fl4})  for $r=2$ under the $\lim_{\p_2,\q_2,c_2\rightarrow 0}$.

The above definitions provide needed foundation to recall on the following fundamental fl-RDT theorem.

\begin{theorem} \cite{Stojnicflrdt23,Stojnicalgbp25,Stojnicalgsbp26}
\label{thm:thmsflrdt1}  Under so-called proportional regime with  $\alpha=\lim_{n\rightarrow\infty} \frac{m}{n}$ remaining constant as  $n$ grows, let elements of $A\in\mR^{m\times n}$ be independent standard normals and let $\cX\subseteq \mS^n$ and $\cY\subseteq \mS^m$ be given sets (subsets of corresponding unit spheres).  Assume that equations (\ref{eq:hmsfl2})--({\ref{eq:fl5}}) hold for an $r\in \mN$ (potentially for $r\rightarrow \infty$ as well). Consider the complete stationarized fl-RDT frame from \cite{Stojnicflrdt23,Stojnicalgbp25} and for a given function $f(\y):R^m\rightarrow R$ set
\begin{align}\label{eq:thmsflrdt2eq1a0}
   \psi_{rp} & \triangleq - \max_{\x\in\cX} s \max_{\y\in\cY} \lp \y^TA\x + f(\y) \rp
   \qquad  \mbox{(\bl{\textbf{random primal}})} \nonumber \\
   \psi_{rd}(\p,\q,\c) & \triangleq    \frac{1}{2}    \sum_{k=2}^{r+1}\Bigg(\Bigg.
   \p_{k-1}\q_{k-1}
   -\p_{k}\q_{k}
  \Bigg.\Bigg)
\c_k
  - \bar{\psi}_{S,\infty}(f(\y),\cX,\cY,\p,\q,\c,s) \quad \mbox{(\bl{\textbf{fl random dual}})}. \nonumber \\
 \end{align}
Let $\hat{\p_0}\rightarrow 1$, $\hat{\q_0}\rightarrow 1$, and $\hat{\c_0}\rightarrow 1$, $\hat{\p}_{r+1}=\hat{\q}_{r+1}=\hat{\c}_{r+1}=0$ and  let the non-fixed parts of $\hat{\p}$, $\hat{\q}$, and  $\hat{\c}$ be the solutions of the following system
\begin{eqnarray}\label{eq:thmsflrdt2eq2a0}
   \frac{d \psi_{rd}(\p,\q,\c)}{d\p} =  0,\quad
   \frac{d \psi_{rd}(\p,\q,\c)}{d\q} =  0,\quad
   \frac{d \psi_{rd}(\p,\q,\c)}{d\c} =  0.
 \end{eqnarray}
 Then,
\begin{eqnarray}\label{eq:thmsflrdt2eq3a0}
    \lim_{n\rightarrow\infty} \frac{\mE_G  \psi_{rp}}{\sqrt{n}}
  & = &
 \lim_{n,r\rightarrow\infty} \psi_{rd}(\hat{\p},\hat{\q},\hat{\c}) \qquad \mbox{(\bl{\textbf{strong sfl random duality}})},\nonumber \\
 \end{eqnarray}
where $\bar{\psi}_{S,\infty}(\cdot)$ is as in (\ref{eq:fl4})-(\ref{eq:fl5}).
 \end{theorem}

The readers familiar with Parisi's RSB ansatz within replica theory \cite{Parisi80,Par79,Par80,Par83} may recognize the nesting averaging of weighted Gaussian sequences. In replicated systems averaging over large number of replicas allows \emph{index decoupling} as a crucial analytical simplification. While such replications are absent in (\ref{eq:hmsfl1})--({\ref{eq:thmsflrdt2eq3a0}), potential simplification when moving from coupled $f_{sq}(\cdot)$ to decoupled $\bar{\psi}_{S,\infty}(\cdot)$ might be useful. Also, the number of nesting levels, $r$, resembles the number of replica breaking steps. However, the mathematical engines developed to achieve the above theorem  rely on nested random processes and relate in no way to any form of replication (for more on the RDT developments, see \cite{StojnicRegRndDlt10,Stojnicflrdt23,Stojnicalgbp25,Stojnicalgsbp26}).

As stated in the preceding paragraph, a simplified structure of the \emph{random dual} is the key advantage that Theorem \ref{thm:thmsflrdt1} offers. However,  even though the results of Theorem \ref{thm:thmsflrdt1} may appear elegant, their ultimate practical significance relies on successful conducting underlying numerical evaluations. Two obstacles often appear on that path: (\textbf{\emph{i}}) In general, sets $\cX$ and $\cY$  may not have component-wise structures which renders their decouplings as questionable; and  (\textbf{\emph{ii}}) Since \emph{a priori}  $r$  can be any positive integer, the convergence in $r$ may be sufficiently slow to prevent concrete numerical evaluations  on higher lifting levels. Both of these obstacles do appear and we will discuss their impact.

\subsection{Fitting negative Hopfield model into Theorem \ref{thm:thmsflrdt1} frame}
\label{sec:neg}

A part of Theorem \ref{thm:thmsflrdt1} power is its generality. Here we need a particular specialization. Set
$\cX= \mB^n $, $\cY=\mS^m$, and
  \begin{eqnarray}\label{eq:prac2}
\varphi(D,\c) & = &
 \mE_{G,{\mathcal U}_{r+1}} \frac{1}{\c_r} \log
\lp \mE_{{\mathcal U}_{r}} \lp \dots \lp \mE_{{\mathcal U}_3}\lp\lp\mE_{{\mathcal U}_2} \lp
\lp    e^{D}   \rp^{\c_2}\rp\rp^{\frac{\c_3}{\c_2}}\rp\rp^{\frac{\c_4}{\c_3}} \dots \rp^{\frac{\c_{r}}{\c_{r-1}}}\rp, \nonumber \\
  \end{eqnarray}
\begin{eqnarray}\label{eq:prac4}
D^{(bin)}(s)
& = &  \sum_{i=1}^n D^{(bin)}_i,  \quad \mbox{with}\quad
D^{(bin)}_i(c_k)=\left |\lp\sum_{k=2}^{r+1}c_k\h_i^{(k)}\rp \right |,
\nonumber \\
   D_i^{(sph)}(b_k) & = &    -\frac{\left \| \sum_{k=2}^{r+1}b_k\u_i^{(2,k)} \right \|^2 }{4\gamma_{sq}} ,
 \end{eqnarray}
and
 \begin{eqnarray}\label{eq:negprac13}
 \lim_{n\rightarrow\infty} \psi_{rd}(\hat{\p},\hat{\q},\hat{\c})
     & = &  \frac{1}{2}    \sum_{k=2}^{r+1}\Bigg(\Bigg.
   \hat{\p}_{k-1}\hat{\q}_{k-1}
   -\hat{\p}_{k}\hat{\q}_{k}
  \Bigg.\Bigg)
\hat{\c}_k
\nonumber \\
& &  - \varphi(D_1^{(bin)}(c_k(\hat{\p},\hat{\q})),\hat{\c}) 
+ \hat{\gamma}_{sq}
- \frac{1}{\alpha}\varphi(-D_1^{(sph)}(b_k(\hat{\p},\hat{\q})),\hat{\c})
\nonumber \\
& \triangleq &   \bar{\psi}_{rd}(\hat{\p},\hat{\q},\hat{\c},\hat{\gamma}_{sq}) .
  \end{eqnarray}

\begin{corollary}
  \label{thme:negthmprac1}
  Assume the setup of Theorem \ref{thm:thmsflrdt1}  with $s=-1$, $f(\y)=0$, $\cX=\mB^n$, and $\cY=\mS^M$. Let $\varphi(\cdot)$ and $\bar{\psi}_{rd}(\cdot)$ be as in (\ref{eq:prac2}) and (\ref{eq:negprac13}), respectively.  Let the ``fixed'' parts of $\hat{\p}$, $\hat{\q}$, and $\hat{\c}$ be such that $\hat{\p}_1\rightarrow 1$, $\hat{\q}_1\rightarrow 1$, $\hat{\c}_1\rightarrow 1$, $\hat{\p}_{r+1}=\hat{\q}_{r+1}=\hat{\c}_{r+1}=0$. Also, let the ``non-fixed'' parts of $\hat{\p}_k$, $\hat{\q}_k$, and $\hat{\c}_k$ ($k\in\{2,3,\dots,r\}$) satisfy
  \begin{eqnarray}\label{eq:negthmprac1eq1}
   \frac{d \bar{\psi}_{rd}(\p,\q,\c,\gamma_{sq})}{d\p} =
   \frac{d \bar{\psi}_{rd}(\p,\q,\c,\gamma_{sq})}{d\q} =
   \frac{d \bar{\psi}_{rd}(\p,\q,\c,\gamma_{sq})}{d\c} =
   \frac{d \bar{\psi}_{rd}(\p,\q,\c,\gamma_{sq})}{d\gamma_{sq}} =  0.
 \end{eqnarray}
One then has
 \begin{eqnarray}
-f_{sq,h}(\infty)
& = &     \frac{1}{2}    \sum_{k=2}^{r+1}\Bigg(\Bigg.
   \hat{\p}_{k-1}\hat{\q}_{k-1}
   -\hat{\p}_{k}\hat{\q}_{k}
  \Bigg.\Bigg)
\hat{\c}_k
  - \varphi(D_1^{(bin)}(c_k(\hat{\p},\hat{\q})),\hat{\c}) 
  +\hat{\gamma}_{sq}
  - \frac{1}{2\alpha}\varphi(-D_1^{(sph)}(b_k(\hat{\p},\hat{\q})),\hat{\c}), \nonumber \\
  \label{eq:negthmprac1eq2}
\end{eqnarray}
with $c_k(\hat{\p},\hat{\q})   =  \sqrt{\hat{\q}_{k-1}-\hat{\q}_k}$ and $b_k(\hat{\p},\hat{\q})   =  \sqrt{\hat{\p}_{k-1}-\hat{\p}_k}$.
\end{corollary}
\begin{proof}
See Appendix \ref{sec:app0}.
\end{proof}

\subsection{Numerical evaluations -- general $\alpha$ and $\kappa$}
\label{sec:numerical}

We utilize Corollary \ref{thme:negthmprac1}  to conduct numerical evaluations. Start from the first lifting level ($r=1$), we incrementally continue to higher levels.

\subsubsection{$r=1$ -- first level of lifting}
\label{sec:firstlev}

For $r=1$ we first note $\hat{\p}_1\rightarrow 1$ and $\hat{\q}_1\rightarrow 1$. Together with $\hat{\p}_{r+1}=\hat{\p}_{2}=\hat{\q}_{r+1}=\hat{\q}_{2}=0$, and $\hat{\c}_{2}\rightarrow 0$  this allows to write
\begin{align}\label{eq:negprac19}
    \bar{\psi}_{rd}^{(1)}(\p,\q,\c,\gamma_{sq})   & =   \frac{1}{2}
\c_2
  - \frac{1}{\c_2}\log\lp \mE_{{\mathcal U}_2} e^{\c_2|\sqrt{1-0}\h_1^{(2)} |}\rp +\gamma_{sq}
- \frac{1}{2\alpha\c_2}\log\lp \mE_{{\mathcal U}_2} e^{-\c_2\frac{\lp \sqrt{1-0}\u_1^{(2,2)} \rp^2}{4\gamma_{sq}}}\rp \nonumber \\
& \rightarrow
  - \frac{1}{\c_2}\log\lp 1+ \mE_{{\mathcal U}_2} \c_2| \h_1^{(2)} |\rp +\gamma_{sq} 
   - \frac{1}{2\alpha\c_2}\log\lp 1- \mE_{{\mathcal U}_2} \c_2\frac{ \lp  \u_1^{(2,2)}  \rp^2 }{4\gamma_{sq}}\rp \nonumber \\
& \rightarrow
   - \frac{1}{\c_2}\log\lp 1+ \c_2\sqrt{\frac{2}{\pi}}\rp +\gamma_{sq}   - \frac{1}{2\alpha \c_2}\log\lp 1- \frac{\c_2}{4\gamma_{sq}}  \rp \nonumber \\
& \rightarrow
  - \sqrt{\frac{2}{\pi}}+\gamma_{sq}
+  \frac{1}{8\alpha\gamma_{sq}}  
 .
  \end{align}
After optimization over $\gamma_{sq}$ one finds the 1-st lifting level estimate for$ f_{sq,h}(\infty)  $
\begin{align}\label{eq:negprac19a0}
 - f_{sq,h}^{(1)}(\infty)  =  \bar{\psi}_{rd}^{(1)}(\p,\q,\c,\gamma_{sq})    
 =
  - \sqrt{\frac{2}{\pi}}+\sqrt{ \frac{1}{2\alpha} }.
  \end{align}
  
The above is then sufficient to numerically evaluate negative Hopfield proxy estimate (obtained on the first level of lifting) for $\kappa_{inf}$
\begin{align}\label{eq:negprac19a1}
\kappa_{inf,-}^{(1)} = - 2\sqrt{2\alpha} f_{sq,h}^{(1)}(\infty)  =  2\sqrt{2\alpha} \bar{\psi}_{rd}^{(1)}(\p,\q,\c,\gamma_{sq})    
 =
  2\sqrt{2\alpha} \lp - \sqrt{\frac{2}{\pi}}+\sqrt{ \frac{1}{2\alpha} } \rp.
  \end{align}
Concrete numerical results are shown in Figure \ref{fig:fig1} as curve denoted by ``Approx Neg Hop 1-spl RDT''. As can be seen, the results are in the vicinity of rigorous lower bounds obtained in previous sections via partially lifted RDT and earlier by Maillard in \cite{Maill24} via first moment analysis. However, one should note that there is a hard limit in the achievable $\alpha$ in low $\kappa$ regime.

\subsubsection{$r=2$ -- second level of lifting}
\label{sec:secondlev}

For $r=2$ we again have $\hat{\p}_1\rightarrow 1$ and $\hat{\q}_1\rightarrow 1$,  and $\hat{\p}_{r+1}=\hat{\p}_{3}=\hat{\q}_{r+1}=\hat{\q}_{3}=0$, but in general   $\hat{\p}_2\neq0$, $\hat{\q}_2\neq0$, and $\hat{\c}_{2}\neq 0$.  Similarly to (\ref{eq:negprac19}), we first write
\begin{eqnarray}\label{eq:negprac24}
    \bar{\psi}_{rd}^{(2)}(\p,\q,\c,\gamma_{sq})   & = &  \frac{1}{2}
(1-\p_2\q_2)\c_2
  - \frac{1}{\c_2}\mE_{{\mathcal U}_3}\log\lp \mE_{{\mathcal U}_2} e^{\c_2|\sqrt{1-\q_2}\h_1^{(2)} +\sqrt{\q_2}\h_1^{(3)} |}\rp \nonumber \\
& &   + \gamma_{sq}
 -\frac{1}{2\alpha\c_2}\mE_{{\mathcal U}_3} \log\lp \mE_{{\mathcal U}_2} e^{-\c_2\frac{\lp  \sqrt{1-\p_2}\u_1^{(2,2)}+\sqrt{\p_2}\u_1^{(2,3)}\rp^2}{4\gamma_{sq}}}\rp.
    \end{eqnarray}
    After solving the integrals, one finds
\begin{eqnarray}\label{eq:negprac24a1}
\zeta_{(z)}^{(2)} = \frac{1}{\c_2} \mE_{{\mathcal U}_3}\log\lp \mE_{{\mathcal U}_2} e^{\c_2|\sqrt{1-\q_2}\h_1^{(2)} +\sqrt{\q_2}\h_1^{(3)} |}\rp
=  \frac{1}{\c_2}  \mE_{{\mathcal U}_3}\log\lp f_{(z)}^{(2)}\rp,
    \end{eqnarray}
where
    \begin{eqnarray}\label{eq:negprac24a0}
f_{(z)}^{(2)} & = & \mE_{{\mathcal U}_2} e^{\c_2|\sqrt{1-\q_2}\h_1^{(2)} +\sqrt{\q_2}\h_1^{(3)} |}
  \nonumber \\
 & = &  
\frac{1}{2} e^{\frac{(1-\q_2)\c_2^2}{2}}
 \Bigg(\Bigg.
 e^{-\c_2\sqrt{\q_2}\h_1^{(3)}}
 \erfc\lp - \lp\c_2\sqrt{1-\q_2}-\frac{\sqrt{\q_2}\h_1^{(3)}}{\sqrt{1-\q_2}}\rp   \rp \nonumber \\
& &  + e^{\c_2\sqrt{\q_2}\h_1^{(3)}}
   \erfc \lp - \lp\c_2\sqrt{1-\q_2}+\frac{\sqrt{\q_2}\h_1^{(3)}}{\sqrt{1-\q_2}}\rp \rp
   \Bigg.\Bigg),
     \end{eqnarray}
and
    \begin{eqnarray}\label{eq:negprac24a0a0}
\zeta_{(w)}^{(2)} & = &\frac{1}{\c_2}\mE_{{\mathcal U}_3} \log\lp \mE_{{\mathcal U}_2} e^{-\c_2\frac{\lp  \sqrt{1-\p_2}\u_1^{(2,2)}+\sqrt{\p_2}\u_1^{(2,3)}\rp^2}{4\gamma_{sq}}}\rp \nonumber \\
& = &  -\frac{1}{\c_2} \log \lp \frac{2\gamma_{sq} + \c_2(1-\p_2)  }{ 2\gamma_{sq}  }  \rp
-
\frac{\p_2}{2\gamma_{sq}(2\gamma_{sq} +\c_2(1-\p_2))}.
     \end{eqnarray}
The above is then enough to get the second level estimate
\begin{eqnarray}\label{eq:negprac24a6}
\kappa_{inf,-}^{(2)} =   -2\sqrt{2\alpha} f_{sq}^{(2,f)}(\infty) & = &     2\sqrt{2\alpha}\bar{\psi}_{rd}^{(2)}(\p,\q,\c,\gamma_{sq}) 
   \rightarrow  
2\sqrt{2\alpha}\lp\frac{1}{2}(1-\p_2\q_2)\c_2
  -  \zeta_{(z)}^{(2)} +\gamma_{sq} - \frac{1}{2\alpha} \zeta_{(w)}^{(2)} \rp.\nonumber \\
    \end{eqnarray}
After taking derivatives and conducting residual numerical evaluations, we obtained the results shown in Figure \ref{fig:fig1}, denoted by ``Approx Neg Hop 2-spl RDT.'' We focused on the second partial level of lifting where $\p_2=\q_2=0$. As shown, the results fit remarkably well between the lower bounds (detailed in the previous section and \cite{Maill24}) and the algorithmic upper bounds of \cite{ChenAlaoui26}. This aligns with what we would expect. Specifically, for the negative Hopfield model, the second partial level of lifting is expected to produce the exact information-theoretic threshold for a range of lower $\alpha_p$ (corresponding to higher $\alpha$ and lower $\kappa$). For higher $\alpha_p$, the optimum is predicted to be on the infinite lifting level with no computational gap, though the difference between the second and higher levels is practically negligible. This implies that for higher $\alpha_p$ (lower $\alpha$), the infinite level may also produce the optimal algorithmic threshold.

Even though the results of \cite{ChenAlaoui26} are not necessarily optimal, their agreement with the RDT prediction is remarkable, even on the second level of lifting. This suggests that minimal modifications to the nonlinearities (or other components of the IAMP variant) used in \cite{ChenAlaoui26} should suffice to achieve the optimal negative Hopfield GSE in the high $\alpha_p$ regime.

\subsubsection{$r$--th level of lifting}
\label{sec:rthlev}

It comes as a no surprise that numerical evaluations become more challenging  as $r$ increases. Efficient $r$-level generalizations are a bit helpful in that regard and we show them next.

Paralleling presentation from the first two levels, for general $r$, we note  $\hat{\p}_1\rightarrow 1$ and $\hat{\q}_1\rightarrow 1$,  and $\hat{\p}_{r+1}=\hat{\q}_{r+1}=0$, and    $\hat{\p}_k\neq0$, $\hat{\q}_k\neq0$, and $\hat{\c}_{k}\neq 0$ for $2\leq k\leq r$.  Analogously to  (\ref{eq:negprac19}) and (\ref{eq:negprac24}), we write
\begin{eqnarray}\label{eq:algalgnegprac24}
    \bar{\psi}_{rd}^{(3)}(\p,\q,\c,\gamma_{sq})   & = &
     \frac{1}{2}
\sum_{k=2}^{r+1}(\p_{k-1}\q_{k-1}-\p_k\q_k)\c_k
 \nonumber \\
& &
  - \frac{1}{\c_r}\mE_{{\mathcal U}_{r+1}}\log \lp \dots \mE_{{\mathcal U}_4}  \lp  \mE_{{\mathcal U}_3}  \lp \mE_{{\mathcal U}_2} e^{\c_2|  \sum_{k=2}^{r+1} c_k\h_1^{(k)}    |}\rp^{\frac{\c_3}{\c_2}}  \rp^{\frac{\c_4}{\c_3}} \dots \rp    \nonumber \\
& &   + \gamma_{sq}
 -\frac{1}{2\alpha\c_r}\mE_{{\mathcal U}_{r+1}} \log \lp \dots \mE_{{\mathcal U}_4}  \lp \mE_{{\mathcal U}_3}   \lp \mE_{{\mathcal U}_2} e^{-\c_2\frac{ \lp   \sum_{k=2}^{r+1} b_k \u_1^{(2,k)} \rp^2}{4\gamma_{sq}}}   \rp^{\frac{\c_3}{\c_2}} \rp^{\frac{\c_4}{\c_3}} \dots \rp ,\nonumber \\
    \end{eqnarray}
where, as in (\ref{eq:fl5}),
\begin{eqnarray}\label{eq:algalgfl5}
  b_k  & = & \sqrt{\p_{k-1}-\p_k} \nonumber \\
c_k   & = & \sqrt{\q_{k-1}-\q_k}.
 \end{eqnarray}
 We set
\begin{eqnarray}\label{eq:algalgnegprac24a2}
\eta_r  =  \sum_{k=3}^{r+1} b_k \u_1^{(2,k)},\quad
\quad
 \theta_r  =  \sum_{k=3}^{r+1} c_k \h_1^{(4)}  . 
\end{eqnarray}
Analogously to (\ref{eq:negprac24a1}) and (\ref{eq:negprac24a0}), we have
\begin{eqnarray}\label{eq:algnegprac24a1}
\zeta_{(z)}^{(r)} = \frac{1}{\c_2} \mE_{{\mathcal U}_3}\log\lp \mE_{{\mathcal U}_2} e^{\c_2|  \sum_{k=2}^{r+1} c_k\h_1^{(k)}    |}   \rp
=  \frac{1}{\c_2}  \mE_{{\mathcal U}_3}\log\lp f_{(z)}^{(r)}\rp,
    \end{eqnarray}
where
    \begin{eqnarray}\label{eq:algnegprac24a0}
f_{(z)}^{(r)} & = & \mE_{{\mathcal U}_2} e^{\c_2|  \sum_{k=2}^{r+1} c_k\h_1^{(k)}    |}  \nonumber \\
 & = &  
\frac{1}{2} e^{\frac{(1-\q_2)\c_2^2}{2}}
 \Bigg(\Bigg.
 e^{-\c_2\sqrt{\q_2}\h_1^{(3)}}
 \erfc\lp - \lp\c_2\sqrt{1-\q_2}-\frac{\theta_r  }  {\sqrt{1-\q_2}}\rp   \rp \nonumber \\
& &  + e^{\c_2\sqrt{\q_2}\h_1^{(3)}}
   \erfc \lp - \lp\c_2\sqrt{1-\q_2}+\frac{\theta_r }  {\sqrt{1-\q_2}}\rp \rp
   \Bigg.\Bigg) .
     \end{eqnarray}
Moreover, after setting
    \begin{eqnarray}\label{eq:algnegprac24a0b0}
\Theta_1 & = & 2\gamma_{sq} \nonumber\\
\Theta_{j} & = & \Theta_{j-1} +\c_j(\p_{j-1}-\p_j) , j=2,3,\dots,r,
\end{eqnarray}
we have analogously to (\ref{eq:negprac24a0a0})
    \begin{eqnarray}\label{eq:algnegprac24a0a0}
\zeta_{(w)}^{(r)} & = &\frac{1}{\c_2}\mE_{{\mathcal U}_3} \log\lp \mE_{{\mathcal U}_2} 
 e^{-\c_2\frac{ \lp   \sum_{k=2}^{r+1} b_k \u_1^{(2,k)} \rp^2}{4\gamma_{sq}}}
  \rp 
  \nonumber \\
& = &  -\frac{1}{\c_2} \log \lp \frac{2\gamma_{sq} + \c_2(1-\p_2)  }{ 2\gamma_{sq}  }  \rp
-
\frac{\p_2}{2\gamma_{sq}(2\gamma_{sq} +\c_2(1-\p_2))}
\nonumber \\
& = &  - \sum_{j=2}^{r}\frac{1}{\c_j} \log \lp \frac{ \Theta_{j} }{ \Theta_{j-1}  }  \rp
-
\frac{\p_r}{2\gamma_{sq}\Theta_r  }
.
     \end{eqnarray}
The above is then enough to get the $r$-th level estimate
\begin{eqnarray}\label{eq:negprac24a6}
\kappa_{inf,-}^{(r)} & = &   -2\sqrt{2\alpha} f_{sq}^{(r)}(\infty)  =      2\sqrt{2\alpha}\bar{\psi}_{rd}^{(r)}(\p,\q,\c,\gamma_{sq}) 
\nonumber \\
&   \rightarrow & 
2\sqrt{2\alpha}\lp \frac{1}{2}\sum_{j=2}^{r}(\p_{j-1}\q_{j-1}-\p_j\q_j)\c_j
  -  \zeta_{(z)}^{(r)} +\gamma_{sq} - \frac{1}{2\alpha} \zeta_{(w)}^{(r)} \rp.
    \end{eqnarray}
The results obtained after taking derivatives and conducting numerical evaluations are shown in Figure \ref{fig:fig1}. We show the results for the 3rd and 4th level of lifting. They are denoted by ``Approx Neg Hop 3-spl RDT'' and  ``Approx Neg Hop 4-spl RDT'', respectively. To be in agreement with the second level, we again focus the partial lifting with $\p_3=\q_3=0$ on the third and $\p_4=\q_4=0$ on the fourth level.

\subsubsection{Discussion}
\label{sec:disc1}

\noindent{\underline{\textbf{\emph{Different $\c$ ordering -- a parametric RDT phenomenon}}}}

As shown in Figure \ref{fig:fig1}, our focus on higher lifting levels centers on the lower $\alpha_p$ (higher $\alpha$) regime. In this regime, a qualitative change in the numerical results occurs starting at the third level. For the presumably natural decreasing order of the $\c$ sequence, we found no further improvement over the second partial level results. However, allowing for arbitrary ordering leads to the emergence of new stationary points, resulting in curves that deviate substantially from the second level curve. We observed a similar phenomenon in our studies of random asymmetric and symmetric binary perceptrons (ABP and SBP) \cite{Stojnicalgbp25,Stojnicalgsbp26}. We formalized this as the main parametric RDT principle, which connects algorithmic thresholds to the potential appearance of statistical-computational gaps.

In contrast, this phenomenon is absent in the higher $\alpha_p$ (lower $\alpha$) regime. At higher levels, we achieve improvements over the second partial level results while maintaining the decreasing order of the $\c$ sequence. (Note: The third and fourth level curves are omitted from the lower $\alpha$ range in Figure \ref{fig:fig1} because the minor visual improvements would overcrowd the figure.)

\vspace{.1in}
\noindent{\underline{\textbf{\emph{Relation with different bounds}}}}

We found one point regarding the behavior in the lower $\alpha$ range particularly interesting. As the improvements with decreasing $\c$ order are possible and no gap in parametric structure is observed, parametric RDT (in agreement with \cite{Stojnicclupsk25}) further predicts the absence of computational gaps for negative Hopfield proxy. Along the same lines, the results from the third and fourth level do improve on those from the second level and bring them even closer to the algorithmic threshold predictions from \cite{ChenAlaoui26}. In the entire range of $\alpha$ where the results from \cite{ChenAlaoui26} are useful ($\alpha \le 0.7178$), our third and fourth level predictions approach them within $\sim 0.01$ (and as $\alpha$ gets lower, within $\sim 0.001$).

Given that our results in these regimes are predicted to be both information-theoretic and algorithmic thresholds, it is remarkably interesting that the approximative I-AMP scheme approaches them so closely. Moreover, the final analytical characterization from \cite{ChenAlaoui26} is remarkably simple. As stated earlier, given such an array of developments, it would not be surprising if very minimal modifications of the algorithmic schemes designed in \cite{ChenAlaoui26} might suffice to achieve optimal negative Hopfield GSE.

Since the curves are almost indistinguishable in the lower $\alpha$ regime, we provide concrete numerical values in Table \ref{tab:tab1}. As can be seen, $\kappa_{inf,-}^{(4)} \approx 1.0748$.

\begin{table}[h]
\caption{$r$-sfl-RDT parameters ($r\leq 4$);
 $\alpha=0.2$; $n\rightarrow\infty$; Union-bounding lower bound $\kappa_{UB}$; Chen - El Alaoui upper bound $\kappa_{CE}$.}\vspace{.1in}
\centering
\def\arraystretch{1.2}
{\small
\begin{tabular}{||l||c||c|c|c||c|c|c||c|c|c||c||}\hline\hline
 \hspace{-0in}$r$-sfl-RDT                                             & $\hat{\gamma}_{sq}$    &  $\hat{\p}_3$ &  $\hat{\p}_2$ & $\hat{\p}_1$     & $\hat{\q}_3$   & $\hat{\q}_2$  & $\hat{\q}_1$ &  $\hat{\c}_4$    & $\hat{\c}_3$    & $\hat{\c}_2$    & $\kappa_{inf,-}^{(r)}$  \\ \hline\hline
$1$-sfl-RDT                                      & $0.7906$ &  $0$  &  $0$  & $\rightarrow 1$   & $0$ &    $0$  & $\rightarrow 1$
 &  $\rightarrow 0$ &  $\rightarrow 0$ &  $\rightarrow 0$  & \bl{$\mathbf{0.9907}$} \\ \hline
   $2$-sfl-RDT                                      & $0.4752$  &   $\rightarrow 0$ & $\rightarrow 0$ & $\rightarrow 1$  &  $\rightarrow 0$  &  $\rightarrow 0$ & $\rightarrow 1$
 &  $\rightarrow 0$ &  $\rightarrow 0$ &  $1.6801$   & \bl{$\mathbf{1.0731}$}  \\ \hline
    $3$-sfl-RDT                                      & $0.4487$  &   $\rightarrow 0$ & $0.8974$ & $\rightarrow 1$  &  $\rightarrow 0$  &  $ 0.7564$ & $\rightarrow 1$
 &  $\rightarrow 0$ &  $1.5098$ &  $2.69$   & \bl{$\mathbf{1.0745}$}  \\ \hline
    $4$-sfl-RDT                                      & $0.4418$  &   $.5719$ & $0.9560$ & $\rightarrow 1$  &  $0.3428$  &  $ 0.88$ & $\rightarrow 1$
 &  $1.2438$ &  $1.7620$ &  $3.5640$   & \bl{$\mathbf{1.0748}$}  \\ \hline
 \hline
  \end{tabular}}
\label{tab:tab1}
\end{table}

From  \cite{ChenAlaoui26} one has the algorithmic upper bound
\begin{equation}
\label{eq:unneghop0a0}
\kappa_{CE} = 2\sqrt{ 1 -2/\pi\sqrt{2\alpha} B + 4/\pi\alpha }, \quad B=2.396280469...
\end{equation}
This bound is shown in Figure \ref{fig:fig1} in its entirety. However, for our considerations here, its practical relevance is in the range $\alpha\leq 0.7178$ or $\kappa\geq 0.5869$. For the very same  $\alpha=0.2$ value as in Table \ref{tab:tab1}, the above bound gives $\kappa_{CE}\approx 1.0767$.

It is also possible to obtain a fairly accurate lower bound utilizing union-bounding concepts. To that end we write
\begin{eqnarray}
\label{eq:unneghop0a1}
\mP\lp 2\sqrt{2\alpha} \min_{\x\in\mB^n}\max_{\y\in\mS^m} \y^TA\x \leq \kappa_{UB} \sqrt{n} \rp
& = &
\mP\lp \min_{\x\in\mB^n}\max_{\y\in\mS^m} \y^TA\x \leq \kappa_c \sqrt{n}\rp
=
\mP\lp \min_{\x\in\mB^n}\|A\x\|_2 \leq \kappa_c \sqrt{n} \rp
\nonumber \\
& = &
\mP\lp \exists \x\in\mB^n \mbox{ such that }\|A\x\|_2 \leq \kappa_c \sqrt{n}\rp
\nonumber \\
& \leq &
2^n\mP\lp  \frac{1}{\sqrt{n}}\|A\1\|_2 \leq \kappa_c \sqrt{n}\rp
\nonumber \\
& = &
2^n\mP\lp \|\g\|_2^2 \leq \kappa_c^2 n\rp
,
\end{eqnarray}
where $\g$ is a vector of $m$ independent standard normals and
\begin{eqnarray}
\label{eq:unneghop0a2}
\kappa_c = \frac{\kappa_{UB}}{2\sqrt{2\alpha} } .
\end{eqnarray}
Utilizing Chernoff bound, we further find for $c>0$
\begin{eqnarray}
\label{eq:unneghop0a3}
 2^n\mP\lp \|\g\|_2^2 \leq \kappa_c^2\rp
\leq 
 2^n  e^{c\kappa_c^2n}  \mE \, e^{ -c\|\g\|_2^2}  
=
 2^n  e^{c\kappa_c^2n}  \lp\frac{1}{\sqrt{1+2c} }\rp^{m}
.
\end{eqnarray}
Combining (\ref{eq:unneghop0a1}) and (\ref{eq:unneghop0a3})
\begin{eqnarray}
\label{eq:unneghop0a4}
\frac{1}{n}\log \lp \mP\lp 2\sqrt{2\alpha} \min_{\x\in\mB^n}\max_{\y\in\mS^m} \y^TA\x \leq \kappa_{UB} \sqrt{n} \rp\rp
&\leq  &\frac{1}{n}
\lp  n\log(2) + c\kappa_c^2n+ m \log \lp\frac{1}{\sqrt{1+2c} }\rp\rp
\nonumber \\
&= & 
\lp  \log(2) + c\kappa_c^2 - \frac{1}{2\alpha} \frac{1}{2}\log \lp 1+2c  \rp\rp.
\end{eqnarray}
The union bound estimate of $\kappa_{inf,-}$ is then $2\sqrt{2\alpha} \kappa_c$, where $\kappa_c$ is such that
\begin{equation}
\label{eq:unneghop0}
\min_{c>0}\lp \log(2) + c\kappa_c^2 - \frac{1}{2\alpha}\frac{1}{2}\log(1+2c) \rp =0.
\end{equation}
Solving over $c$ gives 
\begin{equation}
\label{eq:unneghop1}
c= \frac{1}{2}\lp\frac{1}{2\alpha\kappa_c^2}-1\rp .
\end{equation}
Plugging this back in (\ref{eq:unneghop0}) gives that union-bounding lower bound  $\kappa_{UB}=2\sqrt{2\alpha}\kappa_c$, where $\kappa_c$ is the largest $\kappa\leq \sqrt{\frac{1}{2\alpha}}$ that satisfies 
\begin{equation}
\label{eq:unneghop2}
   \log(2) + \frac{1 -2\alpha \kappa^2 }{4\alpha}  -\frac{1}{4\alpha}\log\lp \frac{1}{2\alpha\kappa^2 } \rp  = 0.
 \end{equation} 
For concreteness, we again take $\alpha=0.2$ (as in Table \ref{tab:tab1}) and find  $\kappa_{UB}=1.0572$. together with the above $\kappa_{CE}$ estimate this gives bouth lower and upper bounds on $\kappa_{inf,-}$
\begin{equation}
\label{eq:unneghop2a0}
\kappa_{UB}\approx 1.0572 < \kappa_{inf,-}^{(4)}\approx 1.0748 < \kappa_{CE}\approx 1.0767.
\end{equation}
 This shows that $\kappa_{CE}$, indeed approaches our predictions within $\sim 0.001$. Moreover, the lower bound obtained via union-bounding arguments is fairly close as well.

\vspace{.1in}
\noindent\underline{\textbf{\emph{Summary:}}}

This is a remarkable development regarding negative Hopfield models. Since \cite{ChenAlaoui26} used them to obtain rigorous algorithmic bounds on matrix discrepancy, one may be tempted to believe that our proxy also maintains an upper bounding nature across the entire $\alpha$ regime. This would suggest that our 2-spl curve might emulate the correct trend of the matrix discrepancy information-theoretic threshold in higher $\alpha$ regimes. More importantly, if the intrinsic features of the negative Hopfield model proxy reflect these information-theoretic properties, they may properly reflect algorithmic trends as well. If so, one might expect a significant computational gap for matrix discrepancy in the higher $\alpha$ regime, though not necessarily precisely where predicted by our theory for the negative Hopfield proxy.

In the following sections, we introduce an algorithmic dynamics that closely follows these predictions in practice. Furthermore, it very closely approaches the theoretical lower bounds in the lower $\alpha$ regime, where the Hopfield proxy is likely far away. This suggests that in such regimes, matrix discrepancy exhibits either a negligible computational gap or no gap at all. This is structurally very similar to the behavior of the negative Hopfield model: union bounding produces lower bounds that closely approach the RDT information-theoretic predictions across the entire range of $\alpha_p$. However, in a certain $\alpha_p$ range, these bounds are far from the algorithmic ones, whereas in other ranges of $\alpha_p$, they are close.

Overall, all of the above suggests the following picture regarding the matrix discrepancy computational gap:

\begin{itemize}
  \item  \emph{Information theoretic $\kappa$ thresholds}: The lower bounds obtained via the first moment analysis in \cite{Maill24} and via lifting in Section \ref{sec:implem} are likely very close to the true values across the entire $\alpha$ range.
 
  \item \emph{Algorithmic $\kappa$ thresholds}: In the lower $\alpha$ range, the gap is either small or nonexistent. In the higher $\alpha$ regime, a significant computational gap may appear. This observation regarding the computational gap in higher $\alpha$ regimes aligns closely with the recent results in \cite{KizLi26}, where stable offline algorithms are proven to achieve no less than $O\lp\sqrt{\frac{1}{\alpha\log(\alpha)}}\rp$, which remains far from the rigorous information-theoretic $O(4^{-\alpha})$ bounds.
  
 \end{itemize}

\vspace{.1in}
\noindent\underline{\textbf{\emph{Analogy with SBP:}}}

The Symmetric Binary Perceptron (SBP) is a vector analogue to the matrix discrepancy problem studied here. While one might expect the two to share information-theoretic and algorithmic properties, whether this is indeed the case remains unknown. However, comparing the matrix discrepancy predictions with those of the SBP provides useful insights.

For the SBP, the information-theoretic threshold is precisely known \cite{AbbLiSly21a,AbbLiSly21b,Aubinetal18,PerkXu21}, and the first moment analysis is exact. In \cite{Stojnicalgsbp26}, the RDT's second lifting level and union bounding produce these same results. Similarly, in Section \ref{sec:implem}, we prove that for matrix discrepancy, the lifted RDT matches the first moment predictions from \cite{Maill24}. Whether these bounds are tight remains an open question. SBP analogy suggests the bound might be tight across the entire $\alpha$ range. Negative Hopfield proxy analogy suggests the bound could be tight in the upper $\alpha$ range, but not necessarily throughout. Alternatively, the bounds from Section \ref{sec:implem} and \cite{Maill24} may be loose across the entire $\alpha$ range. However, our practical algorithmic implementations suggest that any required corrections are likely small, particularly in small $\alpha$ regimes.

When considering algorithmic thresholds, significant differences emerge. A large computational gap is expected in low $\kappa$ regimes for SBP \cite{GamKizPerXu22,BarbAKZ23,Stojnicalgsbp26}. Drawing an analogy to the Negative Hopfield proxy, one might expect a similar gap for matrix discrepancy. Indeed, recent work \cite{KizLi26} proves that for small $\kappa$, fast stable (offline) algorithms for matrix discrepancy cannot drop below $O\lp\frac{1}{-\kappa^2\log\lp\kappa\rp}\rp$. Remarkably, this mirrors the behavior achieved for SBP in \cite{GamKizPerXu22,BarbAKZ23,Stojnicalgsbp26}, where union-bounding provides an exact characterization of the information-theoretic threshold $O(-\log(\kappa))$-matching the scaling union-bounding provides for matrix discrepancy here. Furthermore, a rough small-$\kappa$ analysis of the 3-spl RDT curve in Figure \ref{fig:fig1} yields a similar leading order of $O\lp\frac{1}{\kappa^2 \log^2\lp\kappa\rp}\rp$. Practical algorithmic implementations (discussed below) appear to closely follow this 3-spl curve in smaller $\kappa$ ranges, suggesting that algorithmic behavior for matrix discrepancy is quite similar to that of the SBP. The primary distinction is that for matrix discrepancy, the gap appears significantly smaller in larger $\kappa$ regimes, suggesting it may even be nonexistent in certain portions (as predicted for the negative Hopfield proxy).

 \begin{figure}[h]
\centering
\centerline{\includegraphics[width=1.00\linewidth]{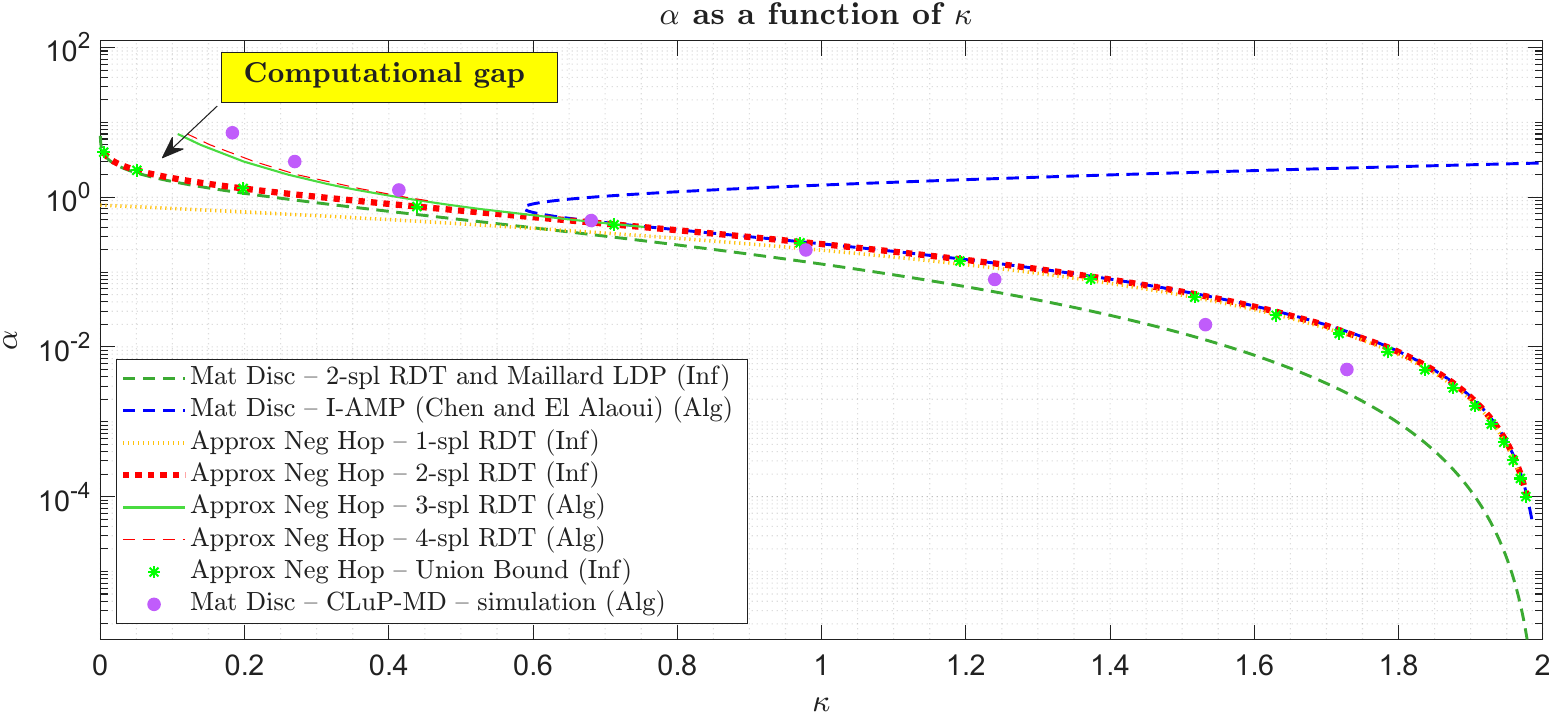}}
\caption{Information theoretic (Inf) and algorithmic (Alg) $\kappa$ thresholds predictions for different $\alpha$; Matrix discrepancy (Mat Disc) and Negative Hopfield proxy (Approx Neg Hop).}
\label{fig:fig1}
\end{figure}

\section{Numerical and algorithmic aspects}
\label{sec:comp}

\subsection{Practicality of numerical evaluations}
\label{sec:pracnumeval}

We find it useful to highlight several aspects related to numerical evaluations. Namely, even when all needed analytical characterizations are available, there is no guarantee that there are efficient numerical procedures to handle them. 

Two primary obstacles appear when solving the stationary points equations numerically: (\textbf{\emph{i}}) increasing lifting level $r$ increases the number of nested integrations which all but assures memory problems; and  (\textbf{\emph{ii}}) opting for iterative schemes more often than not results in procedures being stuck in local optima which coupled with potentially fairly slow convergence could be unsurpassable. Keeping this in mind, we found circumventing  systematic solving approaches and focusing on ad-hoc heuristics as most convenient. Similar strategies were utilized and discussed in detail in \cite{Stojnicalgbp25,Stojnicalgsbp26}. All of those aspect apply here as well. For  convenience, we below briefly reemphasize some of the key points.

As in \cite{Stojnicalgbp25,Stojnicalgsbp26}, we use a semi-manual approach. First a randomly chosen subset of variables is optimized (not necessarily to a full precision). Based on the results of this optimization, other variables are then manually adjusted and a new subset candidates is randomly selected for the subsequent optimization phase. Since it is not clear what the optimal initialization of the numerical procedure should be, we found utilization of the parametric values from the previous lifting level as a helpful strategy. This is particularly beneficial for parameters that appear in earlier levels. However, as each new lifting level introduces new parameters, we found manual trials as a most effective way for their initialization. Avoiding full precision optimizations often allowed preventing the procedure from being stuck in unfavorable local optima. While this is advantageous for navigating rough landscapes, it requires a particular stopping criterion choice. We utilized an overall stopping criterion where the absolute change in the $\kappa$ estimate must be below $10^{-4}$. 

While this ad-hoc heuristic is difficult to formalize, its reliance on manual tryouts provides an opportunity to gain  a bit of intuition regarding the overall flow of the optimization procedure. The approach enables faster navigation through complex landscapes and allows for a better understanding of stationary points. For example, our findings suggest that even if multiple stationary points are present, their associated $\kappa$ values are likely either close to those listed in our table or sufficiently distant to be immediately discarded. Additionally, while some values given in the table may not be fully accurate, the impact of such potential inaccuracies on $\kappa_{inf,-}^{(r)}$ values is expected to be negligible. Interestingly, even significant changes to certain parameters may not noticeably affect $\kappa_{inf,-}^{(r)}$, provided other parameters are adjusted accordingly.

\subsection{CLuP-MD -- a concrete matrix discrepancy algorithm}
\label{sec:algimp}

Under ideal circumstances, the ultimate goal of any matrix discrepancy consideration is to eventually provide an efficient (if possible polynomial) algorithm that for any $\alpha$ outputs $\x\in\mB^n$ with as small $\kappa$ as possible. As the combination of the above analytical machinery and the results available in existing literature suggest that matrix discrepancy likely exhibits a computational gap (at least in some $\alpha$ regimes), approaching information theoretic limits might not be possible. In fact, at present we are unaware of any algorithmic dynamics that practically approach analytical predictions from previous sections. As discussed earlier, the algorithmic phenomenology seems fairly connected to the corresponding ones of SBP and ABP discussed in \cite{Stojnicalgbp25,Stojnicalgsbp26}. Keeping in mind that \cite{Stojnicalgbp25,Stojnicalgsbp26} also introduced CLuP like algorithms with performances well aligned with proposed theoretical predictions, one wonders whether analogous matrix discrepancy designs are possible as well.  In what follows, we introduce a CLuP-MD algorithmic variant which provides a positive answer to this question.

CLuP  (\emph{controlled loosening-up})  algorithms were designed   for solving \emph{planted} models in \cite{Stojnicclupint19}. Considering classical binary and sparse regression problems,  introductory CLuP considerations demonstrated that it achieves high accuracy while avoiding some of the potentially unwanted  features (excessive reliance of planted signal's a priori available knowledge, very large dimensions, and statistical sensitivity) of AMP and other existing alternatives. CLuP variants particularly tailored to fit classical SK models \cite{Stojnicclupsk25} allowed widening the range of applicability to  non-planted scenarios. We show next that a successful CLuP variant tailored for matrix discrepancy can be designed as well.

To that end, we start by noting that matrix discrepancy can be posed as both a feasibility and a standard optimal objective seeking optimization problem. For example, the following would be a direct matrix extension of the feasibility SBP formulation given in (\ref{eq:ex1a0a0})
\begin{eqnarray}
 \mbox{find} & & \x\nonumber \\
\mbox{subject to}
& & \left \|\sum_{i=1}^{n}\bar{A}_i\x_i\right \|_2  \leq \kappa  \nonumber \\
& & \x\in\left \{-\frac{1}{\sqrt{n}}, \frac{1}{\sqrt{n}} \right \}^n\triangleq \mB^n . \label{eq:clupex1a0}
\end{eqnarray}
While this formulation closely resembles similar feasibility formulations from \cite{Stojnicalgbp25,Stojnicalgsbp26}, it is different from the classical optimal objective seeking ones considered  in \cite{Stojnicclupint19,Stojnicclupsk25}. More in alignment with  \cite{Stojnicclupint19,Stojnicclupsk25} is the following variant from (\ref{eq:ex1a1a1})
\begin{eqnarray}
\hspace{-1.5in}\mbox{$\mathbf{\mathcal {MD}}(G,\alpha)$:} \hspace{1in} \kappa = \min_{\x} & & \left \| \sum_{i=1}^{n } \bar{A}_i\x_i \right \|_{2} \nonumber \\
\mbox{subject to}
 & & \x\in \left \{-\frac{1}{\sqrt{n}},\frac{1}{\sqrt{n}} \right \}^n, \label{eq:clupex1}
\end{eqnarray}
To solve (\ref{eq:clupex1}), we propose the following
\begin{eqnarray}\label{eq:algimpeq2}
 \hspace{-.55in} \mbox{\bl{\textbf{\emph{CLuP-MD algorithm:}}}}  \hspace{.65in} \x^{(t+1)} & \rightarrow  &
\mbox{\textbf{gradbar}}\lp f \lp \x; t^{(t)} \rp ;\x^{(t)},t^{(t)} \rp
 \nonumber \\
t^{(t+1)}  &  \rightarrow  & c^{(t)} t^{(t)}.
\end{eqnarray}
 For function $f \lp \x;t^{(t)}  \rp $ specified by argument $t^{(t)}$
\begin{eqnarray}\label{eq:algimpeq3}
f \lp\x;t^{(t)}\rp = - t^{(t)} \|\x\|_2 -  \log\det\lp \kappa_0 I - \sum_{i=1}^{n}\bar{A}_i\x_i \rp
-\frac{1}{n}\sum_{i=1}^{n}  \log(1-n\x_i^2),
\end{eqnarray}
procedure $\mbox{\textbf{gradbar}}\lp f \lp \x; t^{(t)} \rp ;\x^{(t)},t^{(t)} \rp$ applies a form of gradient descent starting from $\x^{(t)}$. A variety of gradient implementations can be used. For example, the one that we used is the following  backtracking line search (parameterized via $\iota_1\in(0,0.5)$ and $\iota_2\in(0,1)$) starting with $t=t_{\iota} =1$
\begin{eqnarray}\label{eq:algimpeq3a0}
\Delta \x & = & -\nabla f \lp\x;t^{(t)}\rp \nonumber \\
t=\iota_2t \quad \mbox{ while } \quad f \lp\x+\Delta \x;t^{(t)}\rp & > & f \lp \x;t^{(t)}\rp + \iota_1 \lp \nabla f \lp\x;t^{(t)}\rp \rp^T\Delta\x \nonumber \\
\x & = &  t\Delta \x.
\end{eqnarray}
The output of the algorithm $\hat{\x}$ is any $\frac{1}{\sqrt{n}}\mathrm{sign}(\x^{(t)})$ that gives the maximum $\kappa$ which we denote $\hat{\kappa}$.

As stated, the above formulation imposes not much restrictions on $\kappa_0$ and  parameters $t^{(0)}$, $c^{(t)}$. Ideal one would prefer a scenario where the objective in (\ref{eq:algimpeq3}) has a favorable  optimizing landscape with as few  local optima as possible. However, achieving such a choice in practice is very delicate and no generic recipe appears to guarantee success (change of ambient dimension $n$ and/or  the matrix size $d$ could be among the reasons). Similarly to \cite{Stojnicalgbp25,Stojnicalgsbp26}, our experience suggests retuning these parameters on fly as best strategy. While overall universality may be unreachable, a few quick pointers might be useful: (\textbf{\emph{i}})  favorable $\kappa_0$ seems to decrease as $\alpha$ increases;  (\textbf{\emph{ii}}) taking $c^{(t)} =1.1$, $t^{(0)}=1.1$, $\iota_1=0.001$, and $\iota_2=0.6$  could serve as a good starting choices set. Also, starting $\x^{(0)}$ is any $\x$ that fits under $\log$s. One can start with a random choice from $\left \{\pm\frac{1}{\sqrt{n}}\right \}^n$ and then scale down by $a_s$ until feasibility is reached. $a_s=1.2$ could serve as a good starting choice. 

We found that restarting for different $\x^{(0)}$ is of critical importance. While increasing the number of restarts improves accuracy, a large number of restarts typically limits the algorithm's applicability in high-dimensional scenarios. However, we were able to apply it to moderate dimensions while keeping the total number of random elements around $10^6$. The obtained results for $\mE\,\hat{\kappa}$ are shown in Figure \ref{fig:fig1}.

To ensure a faithful polynomiality emulation is preserved, the number of restarts never exceeded 200 (a few dozen often sufficed, though increasing the restarts to ~1000 might yield even better results that further support our analytical predictions). We should also note that due to the significant need for retuning, the algorithm may not be practical without further upgrades. Nonetheless, as Figure \ref{fig:fig1} demonstrates, it provides strong support for the analytical considerations presented in the previous section.

\section{Conclusion}
\label{sec:conc}

We have developed the Random Duality Theory (RDT) framework to study random matrix discrepancy in the proportional regime. Utilizing a lifted variant of this framework, we proved lower bounds on discrepancy $\kappa$ that match the corresponding bounds from \cite{Maill24}.

To address the analytical difficulties associated with implementing fully lifted RDT, we propose applying it to a negative Hopfield proxy model. The results obtained for the proxy model at the 2nd lifting level align remarkably well between the information-theoretic lower bounds and the algorithmic upper bounds from \cite{ChenAlaoui26}. At the 3rd and 4th levels of lifting, the parametric RDT concepts predict a sizable lower $\kappa$ computational gap for the proxy model. Extending the information-theoretic  to algorithmic proxy fitting suggests the existence of a similar computational gap for matrix discrepancies in the low $\kappa$ range.

We have also designed a CLuP matrix discrepancy algorithm. Even for dimensions on the order of a few hundreds, its performance closely approaches the information-theoretic predictions in the upper $\kappa$ range and the algorithmic proxy predictions in the lower $\kappa$ range.

\begin{singlespace}
\bibliographystyle{plain}
\bibliography{nflgscompyxRefs}
\end{singlespace}

\appendix

\section{Proof of Corollary \ref{thme:negthmprac1}}
\label{sec:app0}

Keeping in mind  (\ref{eq:prx7}), we note that specializations
$\cX=\left \{-\frac{1}{\sqrt{n}},\frac{1}{\sqrt{n}} \right  \}^n$ and $\cY=\mS^{m}$  allow to rewrite the  \emph{random dual} in the following way
\begin{align}\label{eq:prac1}
    \psi_{rd}(\p,\q,\c) & \triangleq    \frac{1}{2}    \sum_{k=2}^{r+1}\Bigg(\Bigg.
   \p_{k-1}\q_{k-1}
   -\p_{k}\q_{k}
  \Bigg.\Bigg)
\c_k
  - \bar{\psi}_{S,\infty}(0,\cX,\cY,\p,\q,\c,s). \nonumber \\
  & =   \frac{1}{2}    \sum_{k=2}^{r+1}\Bigg(\Bigg.
   \p_{k-1}\q_{k-1}
   -\p_{k}\q_{k}
  \Bigg.\Bigg)
\c_k
  - \frac{1}{n}\varphi(D^{(bin)}(s)) - \frac{1}{n}\varphi(D^{(sph)}(s)), \nonumber \\
  \end{align}
where analogously to (\ref{eq:fl4})-(\ref{eq:fl5})
  \begin{eqnarray}\label{eq:app0prac2}
\varphi(D,\c) & = &
 \mE_{G,{\mathcal U}_{r+1}} \frac{1}{\c_r} \log
\lp \mE_{{\mathcal U}_{r}} \lp \dots \lp \mE_{{\mathcal U}_3}\lp\lp\mE_{{\mathcal U}_2} \lp
\lp    e^{D}   \rp^{\c_2}\rp\rp^{\frac{\c_3}{\c_2}}\rp\rp^{\frac{\c_4}{\c_3}} \dots \rp^{\frac{\c_{r}}{\c_{r-1}}}\rp, \nonumber \\
  \end{eqnarray}
and
\begin{eqnarray}\label{eq:prac3}
D^{(bin)}(s) & = & \max_{\x\in \{-\frac{1}{\sqrt{n}},\frac{1}{\sqrt{n}}\}^n} \lp   s\sqrt{n}      \lp\sum_{k=2}^{r+1}c_k\h^{(k)}\rp^T\x  \rp \nonumber \\
  D^{(sph)}(s) & \triangleq  &    s \max_{\y\in\mS^m}
\lp   \sqrt{n}  \y^T\lp\sum_{k=2}^{r+1}b_k\u^{(2,k)}\rp \rp.
 \end{eqnarray}
Following \cite{Stojnicalgbp25,Stojnicalgsbp26} we set
\begin{eqnarray}\label{eq:prac4}
D^{(bin)}(s)
=  \sum_{i=1}^n D^{(bin)}_i,  \quad \mbox{with}\quad
D^{(bin)}_i(c_k)=\left |\lp\sum_{k=2}^{r+1}c_k\h_i^{(k)}\rp \right |,
\end{eqnarray}
and obtain
  \begin{eqnarray}\label{eq:app0prac6}
\varphi(D^{(bin)}(s),\c) & = &
n \mE_{G,{\mathcal U}_{r+1}} \frac{1}{\c_r} \log
\lp \mE_{{\mathcal U}_{r}} \lp \dots \lp \mE_{{\mathcal U}_3}\lp\lp\mE_{{\mathcal U}_2} \lp
    e^{\c_2D_1^{(bin)}}  \rp\rp^{\frac{\c_3}{\c_2}}\rp\rp^{\frac{\c_4}{\c_3}} \dots \rp^{\frac{\c_{r}}{\c_{r-1}}}\rp
    = n\varphi(D_1^{(bin)}). \nonumber \\
   \end{eqnarray}
We then observe
\begin{eqnarray}\label{eq:prac7}
   D^{(sph)}(s) &  =  &   s \sqrt{n}  \max_{\y\in\mS^m}
\lp      \y^T\lp\sum_{k=2}^{r+1}b_k\u^{(2,k)}\rp \rp
\nonumber \\
& = &    s  \sqrt{n}   \left \| \lp   \sum_{k=2}^{r+1}b_k\u^{(2,k)}\rp  \right \|_2.
 \end{eqnarray}
 Utilizing the  \emph{square root trick}  we further find
\begin{align}\label{eq:prac8}
   D^{(sph)} (s)
& =      s  \sqrt{n}   \left \|  \lp  \sum_{k=2}^{r+1}b_k\u^{(2,k)} \rp  \right \|_2
 =    s\sqrt{n}  \min_{\gamma} \lp \frac{\left \|  \lp  \sum_{k=2}^{r+1}b_k\u^{(2,k)} \rp  \right \|_2^2}{4\gamma}+\gamma \rp 
  \nonumber \\
 & =     s\sqrt{n}  \min_{\gamma} \lp \frac{\sum_{i=1}^{m}   \lp  \sum_{k=2}^{r+1}b_k\u_i^{(2,k)}\rp ^2}{4\gamma}+\gamma \rp.
 \end{align}
 Scaling  $\gamma_{sq}=\gamma_{sq}\sqrt{n}$  and specialization $s=-1$ give
\begin{eqnarray}\label{eq:prac9}
   D^{(sph)}(s)
  & =  &     -\sqrt{n}  \min_{\gamma_{sq}} \lp \frac{\sum_{i=1}^{m} \lp  \sum_{k=2}^{r+1}b_k\u_i^{(2,k)} \rp^2}{4\gamma_{sq}\sqrt{n}}+\gamma_{sq}\sqrt{n} \rp   \nonumber \\
  & = &   - \min_{\gamma_{sq}} \lp \frac{\sum_{i=1}^{m}  \lp  \sum_{k=2}^{r+1}b_k\u_i^{(2,k)}  \rp^2}{4\gamma_{sq}}+\gamma_{sq}n \rp  
  =    - \min_{\gamma_{sq}} \lp \sum_{i=1}^{m} D_i^{(sph)}(b_k)+\gamma_{sq}n \rp, 
 \end{eqnarray}
with
\begin{eqnarray}\label{eq:app0prac10}
   D_i^{(sph)}(b_k)= \frac{\lp \sum_{k=2}^{r+1}b_k\u_i^{(2,k)} \rp^2}{4\gamma_{sq}}.
 \end{eqnarray}
After noting the following connection between the GSE, $f_{sq}(\infty)$, from (\ref{eq:frlimlogpartfunsqrta0}), and the random primal, $\psi_{rp}(\cdot)$, from Theorem \ref{thm:thmsflrdt1},
 \begin{equation}
-f_{sq,h}(\infty)
    = \lim_{n\rightarrow\infty} \frac{\mE_G  \psi_{rp}}{\sqrt{n}}
   =
 \lim_{n\rightarrow\infty} \psi_{rd}(\hat{\p},\hat{\q},\hat{\c}),
  \label{eq:negprac11}
\end{equation}
 one obtains through a combination of (\ref{eq:prac1})-(\ref{eq:negprac11})
 \begin{eqnarray}\label{eq:app0negprac13}
 \lim_{n\rightarrow\infty} \psi_{rd}(\hat{\p},\hat{\q},\hat{\c})
     & = &  \frac{1}{2}    \sum_{k=2}^{r+1}\Bigg(\Bigg.
   \hat{\p}_{k-1}\hat{\q}_{k-1}
   -\hat{\p}_{k}\hat{\q}_{k}
  \Bigg.\Bigg)
\hat{\c}_k
\nonumber \\
& &  - \varphi(D_1^{(bin)}(c_k(\hat{\p},\hat{\q})),\hat{\c}) + \hat{\gamma}_{sq}- \frac{1}{2\alpha}\varphi(-D_1^{(sph)}(b_k(\hat{\p},\hat{\q})),\hat{\c})
\nonumber \\
& \triangleq &   \bar{\psi}_{rd}(\hat{\p},\hat{\q},\hat{\c},\hat{\gamma}_{sq}) .
  \end{eqnarray}
From  (\ref{eq:negprac11}) and (\ref{eq:negprac13}) we also have
 \begin{eqnarray}
-f_{sq,h}(\infty)
  & = &
 \lim_{n\rightarrow\infty} \psi_{rd}(\hat{\p},\hat{\q},\hat{\c})
 =   \bar{\psi}_{rd}(\hat{\p},\hat{\q},\hat{\c},\hat{\gamma}_{sq}) \nonumber \\
 & = &   \frac{1}{2}    \sum_{k=2}^{r+1}\Bigg(\Bigg.
   \hat{\p}_{k-1}\hat{\q}_{k-1}
   -\hat{\p}_{k}\hat{\q}_{k}
  \Bigg.\Bigg)
\hat{\c}_k
  - \varphi(D_1^{(bin)}(c_k(\hat{\p},\hat{\q})),\hat{\c}) + \hat{\gamma}_{sq} - \frac{1}{2\alpha} \varphi(-D_1^{(sph)}(b_k(\hat{\p},\hat{\q})),\hat{\c}). \nonumber \\
  \label{eq:negprac18}
\end{eqnarray}
The preceding discussion together with Theorem \ref{thm:thmsflrdt1} is then sufficient to complete the proof.

\end{document}